\documentclass[]{aastex631}

\newcommand{\arcsinh}{\text{arcsinh}}

\shorttitle{Neutron Stars with Dark Matter Halos}
\shortauthors{Shawqi \& Morsink}

\begin{document}

\title{Interpreting Mass and Radius Measurements of Neutron Stars with Dark Matter Halos}

\email{shawqi@ualberta.ca, morsink@ualberta.ca}

\author[0000-0002-1095-6183]{Shafayat Shawqi}
\affiliation{Department of Physics, University of Alberta, Edmonton, AB, T6G 2E1, Canada}

\author[0000-0003-4357-0575]{Sharon M. Morsink}
\affiliation{Department of Physics, University of Alberta, Edmonton, AB, T6G 2E1, Canada}



\begin{abstract}

The high densities of neutron stars (NSs) could provide astrophysical locations for dark matter (DM) to accumulate. Depending on the DM model, these DM admixed NSs (DANSs) could have significantly different properties than pure baryonic NSs, accessible through X-ray observations of rotation-powered pulsars. We adopt the two-fluid formalism in general relativity to numerically simulate stable configurations of DANSs, assuming a fermionic equation of state (EOS) for the DM with repulsive self-interaction. The distribution of DM in the DANS as a halo affects the path of X-rays emitted from hot spots on the visible baryonic surface causing notable changes in the pulse profile observed by telescopes such as NICER, compared to pure baryonic NSs. We explore how various DM models affect the DM mass distribution, leading to different types of dark halos. We quantify the deviation in observed X-ray flux from stars with each of these halos.   We identify the pitfalls in interpreting mass and radius measurements of NSs inferred from electromagnetic radiation and constraining the baryonic matter EOS, if these dark halos exist.

\end{abstract}



\section{Introduction} \label{sec:intro}

If dark matter (DM) does not self-annihilate, it can accumulate inside stars such as the Sun \citep{Press1985}, other main sequence stars {\citep{Kouvaris2011, Iocco2012}}, white dwarf stars \citep{Bramante2015}, and neutron stars (NSs) \citep{Goldman1989}. DM admixed NSs (DANSs) are of particular interest since the high densities in their cores and high temperatures reached during their births could lead to a significant production of DM particles from standard model particles \citep{Ellis2018,  Nelson2019}. 

Adding DM to a NS changes the distribution of matter, leading to changes in the mass, radius, and tidal deformability. Since these properties can be measured through observations of electromagnetic and gravitational radiation, this leads to the possibility that observations of DANSs could constrain the properties of DM. However, the equation of state (EOS) of dense baryonic matter (BM) is unknown. Since different choices of a BM-EOS also affect the possible masses and radii of NSs, the properties of both DM and BM must be considered simultaneously. 

One method for constraining the properties of DANSs is through the observations of gravitational radiation emitted by merging NSs. Since a NS (or a DANS) is not a point particle, it will be tidally deformed by its companion, leading to observable changes in the gravitational radiation's phase compared to point-particle waveforms. Depending on the assumed properties of the DM, the DANS could have a smaller tidal deformability 
\citep{Ellis2018} or a larger deformability \citep{Nelson2019}. Limits on the tidal deformability from LIGO observations \citep{Abbot2017} have been used by several authors to place some limits on the properties of DM (see \cite{Nelson2019, Karkevandi2022, Mariani2024} for examples).


X-rays emitted from the surface of a NS are gravitationally lensed by the NS's strong gravitational field. Observations of this light are used to constrain the masses and radii of NSs, either through observations of pulsed X-rays by the Neutron star Interior Composition ExploreR (NICER) \citep{Riley2019,Miller2019}, or through unpulsed emission by Chandra or other X-ray telescopes \citep{Steiner2018}. \cite{Miao2022} and \cite{Shakeri2024} showed that the existence of dark halos (fermionic and bosonic, respectively) can significantly alter the X-ray pulse profiles observed by telescopes like NICER. \cite{Rutherford2023} investigated the capabilities of current and future X-ray telescopes to constrain DM properties in the case of dark cores. Separate from constraining DM parameters, a major goal of electromagnetic and gravitational wave astronomy missions observing NSs has been to derive constraints on the EOS of dense BM. If significant amounts of DM exist in NSs, these constraints would not be accurate. In this paper, we explore in more detail many issues related to the interpretation of light emitted from the surface of a DANS, as opposed to attempting to constrain the microscopic properties of DM.

While it is unlikely that DM capture over the entire lifetimes of NSs from the galactic halo is too inefficient to significantly affect their mass-radius relation \citep{Goldman1989, Kouvaris2008, Kouvaris2010, Nelson2019}, the extreme internal density and temperature environments at different stages of their lifetime can allow production mechanisms for sizable amounts of DM (up to a few percents of the total star mass) from standard model species. As a possible solution to the neutron decay anomaly \citep{Paul2009}
it was proposed by \cite{Fornal2018} that free neutrons may decay through a dark channel $\left( n \to \chi + \dots \right)$. While systematic and statistical uncertainties could be the cause of the  anomaly \citep{Czarnecki2018, Dubbers2019}, the dark decay channel of the neutron has introduced a pathway for DM to be produced in the extremely high neutron-dense cores of NSs \citep{Motta2018a, Ellis2018, Baym2018, McKeen2018, Motta2018b}. Alternatively, high temperatures during progenitor supernovae and in proto-NSs allow DM production through the neutron Bremsstrahlung process $\left( nn \to nn\chi \right)$ \citep{Ellis2018, Nelson2019, Reddy2022}. Yet another formation mechanism for these DM admixed NSs (DANSs) requires the existence of the hypothesized dark stars \citep{Kouvaris2015}. 
If DM self-interaction exists, dark stars could form which provide a number of formation pathways for DANSs: (i) pure baryonic NSs could merge with dark stars \citep{Ciarcelluti2011}, (ii) dark stars could act as gravitational seeds for baryonic star formation, which eventually turn into DANSs \citep{Kamenetskaia2022}.

In this paper, we assume that some amount of asymmetric (non-self-annihilating) DM exists gravitationally bound to the NSs in non-rotating stable equilibrium configurations and explore the mass-radius (MR) relation and gravitational self-lensing properties of the objects. We use the two-fluid formalism in general relativity to numerically simulate the DANSs. The two-fluid formalism provides a natural way to distinguish the baryonic radius, $R_B$, where light is emitted, and the dark radius, $R_D$, the outer edge of the region containing dark matter. In this formalism, these objects can be distinctly classified into \textit{dark cores} or \textit{dark halos}, based on the distribution of the dark and baryonic matter in the DANS. In the case of a dark core, all the DM exists within the baryonic radius of the star $\left( R_{B} \geq R_{D} \right)$. In the case of a dark halo, the DM extends throughout the star and beyond $R_{B}$ to form a DM halo $\left( R_{D} > R_{B} \right)$. For a particular pure baryonic NS, whether a dark core or halo forms when DM is added to it, depends on (i) the bosonic or fermionic nature of the DM, (ii) the DM particle mass $\left( m_{\chi} \right)$, (iii) the DM self-interaction strength $\left( y \right)$, and (iv) the total DM mass as a fraction of the total mass of the DANS $\left( f_{\chi} \right)$. These properties are the contributing factors to the DM enthalpy. A dark core forms when the central enthalpy of the baryonic matter (BM) is larger than or equal to that of the DM \citep{Miao2022}. Conversely, if the central enthalpy of the DM is larger than that of the BM, a dark halo forms. We choose the DM to be fermionic and explore properties with a wide range of repulsive self-interaction strengths including zero.

The rest of the paper is organized as follows. In Section \ref{sec:model} we set up the theoretical framework to model DANSs as two fluids in hydrostatic equilibrium in general relativity using the TOV equations. We also describe the distribution of the baryonic and dark matter in these stars and classify dark halos as either \textit{compact} or \textit{diffuse}, based on the fraction of its dark matter mass residing in or outside  $R_B$. In Section \ref{sec:MR}, we compute MR curves for NSs with self-interacting fermionic dark matter and two different BM-EOSs. The qualitative properties of \textit{compact} and \textit{diffuse} halos evident on their MR curves are discussed. In Section \ref{sec:lens}, we provide detailed descriptions of the gravitational self-lensing properties of DANSs with dark halos and show how the no-halo Schwarzschild counterpart approximation (introduced by \citep{Miao2022}) arises as a limiting behavior of the lensing equations. In Section \ref{sec:methods} we discuss the different types of observations and how they can be used in EOS inference work.
We use Planck units with $c = G = \hbar = 1$. 

\section{DANS Model Generation} \label{sec:model}

DANSs can be mathematically modeled as a single fluid or two separate non-interacting fluids. In the single{-}fluid formalism, the DM couples to the BM through Standard Model mediators along with gravity. A combined EOS of all the constituents is then calculated and equilibrium configurations are obtained by solving the Tolman-Oppenheimer-Volkoff (TOV) equations \citep{Tolman1939, Oppenheimer1939}. While this is a perfectly valid method to employ in modeling DANSs, 
it may not be ideal to analyse the emission of light from DANSs. This is because the definition of the baryonic surface is not as straight-forward in the single-fluid description. Since NSs emit X-rays from hot spots on their baryonic surface, it is crucial to determine the baryonic radius. This allows the calculation of how much extra light-bending is caused by the DM situated outside $R_{B}$.

By keeping track of the number of baryons in the single fluid method, it may be possible to assign the baryonic surface to the location where the number of baryons go to zero. We leave this investigation for the future, and work within the simpler two-fluid formalism in this paper.

\subsection{Two-fluid Formalism}

In the two-fluid formalism, it is assumed that the Standard Model couplings between the BM and DM are negligible, and the two fluids only interact through gravity. The bulk properties of the two fluids are accounted for separately so it is possible to clearly define $R_{B}$ and $R_{D}$. 

The metric for a spherically symmetric spacetime is
\begin{equation}
    ds^2 = - e^{2 \Phi(r)} dt^2 + e^{2 \Lambda(r)} dr^2 + r^2d\Omega^2.
    \label{eq:metric}
\end{equation}
The two-fluid TOV equations \citep{Kodama1972, Sandin2009} are
\begin{eqnarray}
e^{-2\Lambda(r)} &=& 1 - \frac{2 M_T(r)}{r}, \label{eq:lambda} \\
\frac{d\Phi \left( r \right)}{dr} &=& \frac{ M_{T} \left( r \right) + 4 \pi r^{3} P \left( r \right) }{r^{2} e^{-2\Lambda(r)}}, \label{eq:dphidr} \\
\frac{dP_{B} \left( r \right)}{dr} &=& - \left[ \epsilon_{B} \left( r \right) + P_{B} \left( r \right) \right] \frac{d\Phi \left( r \right)}{dr}, \\
\frac{dP_{D} \left( r \right)}{dr} &=& - \left[ \epsilon_{D} \left( r \right) + P_{D} \left( r \right) \right] \frac{d\Phi \left( r \right)}{dr}, \\
\frac{dM_{B} \left( r \right)}{dr} &=& 4 \pi r^{2} \epsilon_{B} \left( r \right), \\
\frac{dM_{D} \left( r \right)}{dr} &=& 4 \pi r^{2} \epsilon_{D} \left( r \right), \label{eq:dM_Ddr}
\end{eqnarray}
where $M_{T} \left( r \right) = M_{B} \left( r \right) + M_{D} \left( r \right)$, $P \left( r \right) = P_{B} \left( r \right) + P_{D} \left( r \right)$, $\epsilon \left( r \right) = \epsilon_{B} \left( r \right) + \epsilon_{D} \left( r \right)$, and the subscripts $B$ and $D$ represents baryonic and dark matter, respectively. The gravitational redshift, $z(r)$ between light emitted at $r$ and detected by an observer at infinity is defined by $1+z(r) = e^{-\Phi(r)}$. In particular, we will be concerned with light emitted from the baryonic surface so we define $z_B=z(R_B)$ and $\Phi_B = \Phi(R_B)$.

One EOS each, relating pressure and energy density, for the baryonic and the dark matter is then required to generate equilibrium configurations of DANSs.  We use the stiff BM-EOS, NL3$\omega \rho$L55 \citep{Horowitz2001, Pais2016}, retrieved from the CompOSE online repository of EOSs \citep{Typel2015, Oertel2017, CompOSE2022}. 
To contrast with the stiff BM-EOS, we provide some results with the softer BM-EOS, SLy5 \citep{Chabanat1998,Grams2022}.
We do not expect the qualitative aspects of our result to change based on the choice of BM or DM EOS.
CompOSE provides tabulated values of thermodynamic properties for the BM-EOSs. We interpolate between the minimum and maximum provided values of the functions of interest using 500 grid points with logarithmic scaling and requiring the quantities to be continuous in function values.

We adopt a self-interacting fermionic DM-EOS, however, similar conclusions would be found if a bosonic DM-EOS were used, as done by \cite{Shakeri2024}. 
The EOS for self-interacting fermionic DM is
\citep{Narain2006,Nelson2019}
\begin{eqnarray}
\epsilon_{D} = \frac{m_{\chi}^{4}}{8 \pi^{2}} \left[ \left( 2x^{3} + x \right) \sqrt{1 + x^{2}} - \arcsinh{\left( x \right)} \right] + \frac{m_{\chi}^{4} y^{2} x^{6}}{\left( 3 \pi^{2} \right)^{2}}, \\
P_{D} = \frac{m_{\chi}^{4}}{24 \pi^{2}} \left[ \left( 2x^{3} - 3x \right) \sqrt{1 + x^{2}} + 3 \arcsinh{\left( x \right)} \right] + \frac{m_{\chi}^{4} y^{2} x^{6}}{\left( 3 \pi^{2} \right)^{2}},
\end{eqnarray}
where $m_\chi$ is the mass of the DM particle and $y$ is the self-interaction strength, defined by $y=m_{\chi}/m_I$ where $m_I$ is the mass of the self-interaction force mediator. The relativity parameter, $x$ is defined by $x=k_F/m_\chi$, where $k_F$ is the DM Fermi momentum. 

Assigning a particular pair of baryonic and dark matter central energy densities for each DANS model, the two-fluid TOV equations are numerically solved using the fourth-order Runge-Kutta-Fehlberg (RKF4(5)) adaptive step-size method. The equations are integrated radially outwards with the pressure and energy density of each fluid evaluated at each step. The baryonic and dark radii, $R_{B}$ and $R_{D}$, respectively, are assigned to the corresponding radial steps where the enclosed mass derivative of each of the fluids goes to zero. The gravitational masses of each of the fluids are $M_B = M_{B} \left( R_{B} \right)$ and $M_D = M_{D} \left( R_{D} \right)$, with $M_{T} = M_{B} + M_{D}$, the total gravitational mass of the DANS. The total DM mass as a fraction of the total mass of the DANS is then $f_{\chi}\footnote{\textrm{Not to be confused with $f_{\chi}$ as defined by \cite{Mariani2024} to be the central pressure of the DM fluid as a fraction of the total pressure at the center of the star, i.e. $P_{D} \left( 0 \right) / P \left( 0 \right)$.}} = M_{D}/M_{T}$. While it is common practice, in codes using the RK4 method, to assign the NS surface to the radial step where pressure vanishes, we find that our method provides more accurate results when using adaptive step sizes.

\subsection{Distribution of Mass in a DANS}

NSs emit X-rays from hot regions on their baryonic surface. So, to understand their self-lensing it is important to keep track of the total enclosed gravitational mass at $R_B$, i.e. $M_{T} \left( R_{B} \right) = M_{B} \left( R_{B} \right) + M_{D} \left( R_{B} \right)$. In the case of a dark core, $M_{T} \left( R_{B} \right) = M_{T}$  while $M_{T} \left( R_{B} \right) < M_{T}$ for dark halos. Thus, for comparing self-lensing properties between stars with dark halos and those without, we separate the DM mass distribution in halos into two parts, namely the DM mass located inside $R_B$, $M_{D} \left( R_{B} \right)$, and the DM mass located outside $R_B$ (the DM cloud\footnote{This quantity is denoted $M_{\textrm{halo}}$ by \cite{Miao2022}. However, in galactic DM terminology, the DM halo mass corresponds to all of the DM both inside {\em{and}} outside of the visible part of the galaxy. To reduce confusion, we prefer not to use the word halo to refer to the region outside of the visible DANS surface.}), $M_{c} = M_{T} - M_{T} \left( R_{B} \right)$.
To illustrate the effect of a fairly large dark matter fraction, we adopt $f_\chi=0.05$, as suggested by formation scenarios introduced by \citet{Ellis2018}.


\begin{deluxetable}{lccccccccccc}[ht!]
  \tabletypesize{\scriptsize} 
  \tablecolumns{11}
  \tablewidth{0pt}
  \tablecaption{
    Properties of DANSs of mass $M_{T} = 1.4 M_{\odot}$ and $f_{\chi} = 0.05$ with BM-EOS NL3$\omega \rho$L55 and varying values of $y$ and $m_{\chi}$, 
    and two reference NS constructed with the same BM-EOS. Color names refer to the colors used for the different DM-EOS in the figures.
    The mass contained in the cloud outside the baryonic surface is denoted $M_c$. 
    \label{tab:MRproperties}
  }
  \tablehead{%
    \colhead{} &
    \colhead{} &
    \colhead{$m_{\chi} $} &
    \colhead{$y$} &
    \colhead{$M_{T} $} &
    \colhead{$M_{T} \left( R_{B} \right) $} &
    \colhead{$M_{\textrm{c}} $} &
    \colhead{$R_{B} $} &
    \colhead{$R_{D} $} &
    \colhead{$M_{\textrm{c}}/R_{D}$} &
    \colhead{$M_{\textrm{c}}/M_{D}$}
    \\
    \colhead{} &
    \colhead{} &
    \colhead{$\left[ \textrm{GeV} \right]$} &
    \colhead{} &
    \colhead{$\left[ M_{\odot} \right]$} &
    \colhead{$\left[ M_{\odot} \right]$} &
    \colhead{$\left[ M_{\odot} \right]$} &
    \colhead{$\left[ \textrm{km} \right]$} &
    \colhead{$\left[ \textrm{km} \right]$} &
    \colhead{$\left[ 10^{-3} \right]$} &
    \colhead{}
  }

  \startdata
    pure NS & black & $0$ & $0$ & $1.40$ & $1.40$ & $0$ & $13.76$ & $0$ & N/A & N/A \\
    pure NS & black & $0$ & $0$ & $1.33$ & $1.33$ & $0$ & $13.71$ & $0$ & N/A & N/A \\
    \hline
    dark core & green & $1$ & $0$ & $1.40$ & $1.40$ & $0$ & $13.35$ & $6.48$ & $0$ & $0$ \\
    compact halo & yellow & $0.3$ & $0$ & $1.40$ & $1.39$ & $0.0087$ & $13.47$ & $19.38$ & $0.66$ & $0.124$ \\
    diffuse halo & pink & $0.15$ & $0$ & $1.40$ & $1.34$ & $0.0599$ & $13.67$ & $122.75$ & $0.72$ & $0.853$ \\
    diffuse halo & sky blue & $0.1$ & $0$ & $1.40$ & $1.33$ & $0.0678$ & $13.70$ & $399.41$ & $0.25$ & $0.969$ \\
    \hline
    compact halo & orange & $1$ & $25$ & $1.40$ & $1.40$ & $0.0003$ & $13.44$ & $13.93$ & $0.035$ & $0.005$ \\
    intermediate halo & red & $1$ & $100$ & $1.40$ & $1.34$ & $0.0557$ & $13.66$ & $48.54$ & $1.7$ & $0.793$ \\
    diffuse halo & blue & $1$ & $1000$ & $1.40$ & $1.33$ & $0.0702$ & $13.71$ & $526.41$ & $0.20$ & $0.997$
\enddata
\end{deluxetable}

We have computed the structure of a wide range of DANS models using a self-interacting fermionic DM model and the NL3$\omega \rho$L55 baryonic EOS. The values of the baryonic and dark radii $R_B$ and $R_D$ are shown in Table \ref{tab:MRproperties} for a few representative DANS models 
with $M_T=1.4 M_\odot$ and $f_\chi = 0.05$, corresponding to a total baryonic mass of $M_B = 1.33 M_{\odot}$ and DM mass of $M_D = 0.07 M_\odot$. For each DANS, we also list the values of two dimensionless ratios. The ratio of $M_c/R_D$, a diagnostic ratio introduced by \citep{Miao2022}, is proportional to the difference in the metric at $R_D$ compared to the metric of a reference NS with a mass equal to $M_T(R_B)$. The ratio $M_c/M_D$ indicates whether the dark matter is mainly outside or inside the baryonic surface of the DANS.

\begin{figure}[ht!]
\plotone{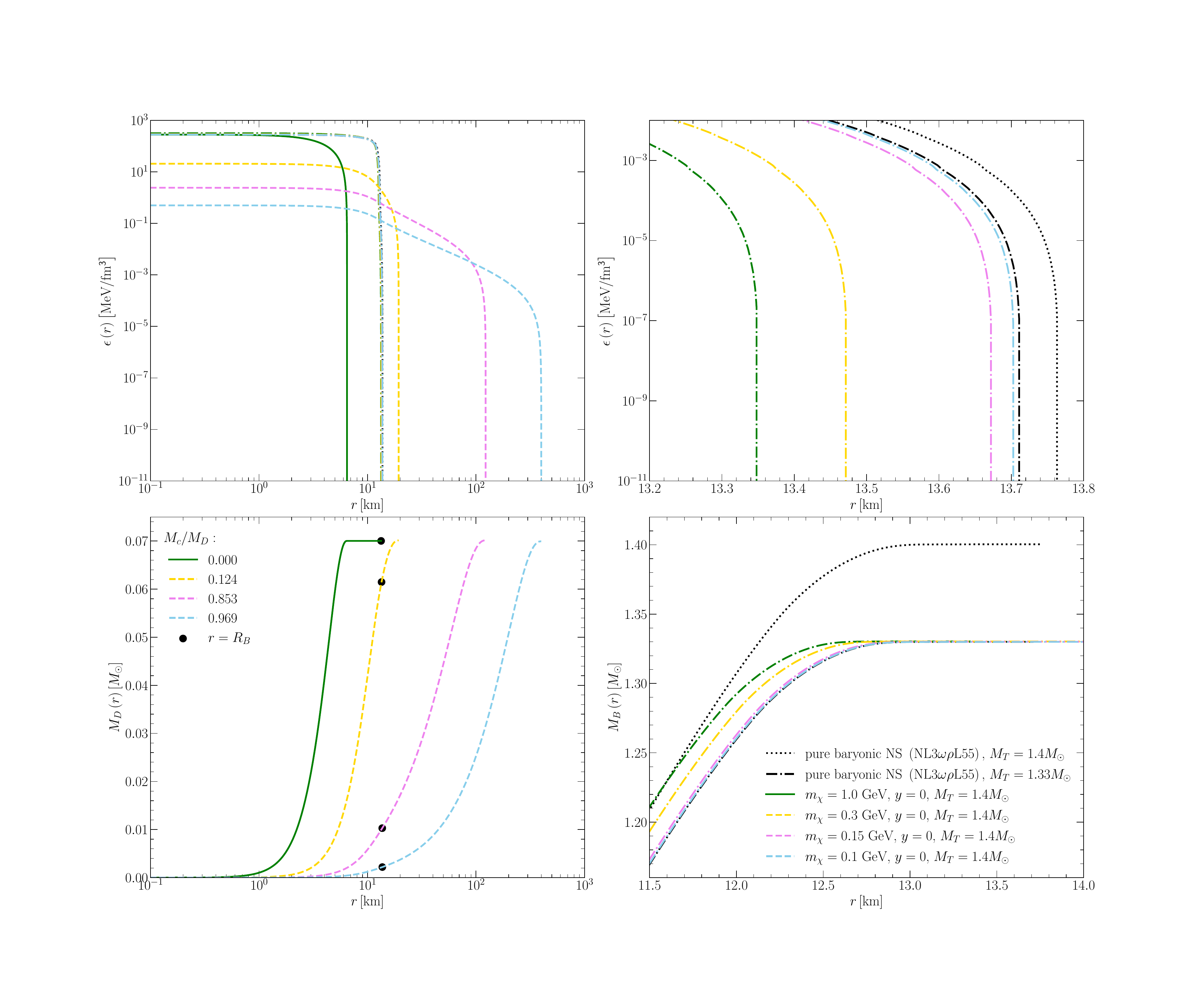}
\caption{Density and mass distributions for DANSs with varying DM particle mass $m_{\chi}$, self-interaction parameter $y = 0$, and DM fraction $f_{\chi} = 0.05$, and baryonic EOS NL3$\omega \rho$L55. Top left: Baryonic (dash-dot) and dark matter (solid or dashed) energy densities as a function of radial distance from the centre of $M_{T} = 1.4 M_{\odot}$ DANS models. Top right: Baryonic (dash-dot) energy densities for the same DANS models. Energy densities of a $M_{T} = 1.33 M_{\odot}$ (black dash-dot) and a  $M_{T} = 1.4 M_{\odot}$ (black dotted) pure baryonic NSs are also plotted for comparison. Bottom left: Enclosed DM mass as a function of radial distance. Black dots show the location of the baryonic surface of the DANS. The mass fraction of DM in the cloud is shown in the legend. Bottom right: Enclosed baryonic matter mass as a function of radial distance in the region close to the baryonic surface.
\label{fig:density-yconst}}
\end{figure}

The distribution of energy density and mass is shown in Figure~\ref{fig:density-yconst} for the first four DANS models shown in Table~\ref{tab:MRproperties}. 
Figure \ref{fig:density-yconst} (top left) shows the baryonic (dash-dot) and dark matter (solid or dashed) energy density profiles. The green solid curve represents a dark core and the dashed curves represent dark halos. For large $m_{\chi}$, the DM tends to accumulate near the centre of the star, forming a dense dark core. As $m_{\chi}$ is reduced, the dark fluid is allowed to get puffier, pushing the dark radius, $R_{D}$, outwards beyond $R_{B}$ to form larger and larger halos. The logarithmic scale used in this panel makes the 6 baryonic energy density curves appear to overlap.  Figure \ref{fig:density-yconst} (top right) shows the baryonic (dash-dot) energy densities of the same DANS models using a linear scale. 
Two reference NSs are shown, one with $M=1.4 M_\odot$, to compare with $M_T$ and another with $M = 1.33 M_\odot$ to compare with the baryonic mass of each of the DANS models.
Due to the additional gravitational attraction of the DM within $R_{B}$, the baryonic radius is pulled inwards in the 4 DANS models compared to the two reference pure baryonic NSs (black). The larger $m_{\chi}$ is, the larger the DM mass within $R_{B}$, and thus the more $R_{B}$ is pulled inwards. It should be emphasized that the effect of reducing the visible radius of the DANS is not limited to dark cores, as more commonly studied in the literature. While the effect is maximized in the case of cores, sufficiently small repulsive self-interactions 
can shrink $R_B$ in the case of halos too, as noted by \citet{Shakeri2024} for DANSs with bosonic DM.

Figure \ref{fig:density-yconst} (bottom) shows the enclosed dark (left) and baryonic (right) matter mass distributions for the same DANSs. The amount of DM mass situated within $R_{B}$ increases with increasing $m_{\chi}$, as is shown in Table \ref{tab:MRproperties} for each of these models. 
The baryonic energy density of the $m_{\chi} = 0.1 \textrm{ GeV}$ case closely follows that of the $M_{T} = 1.33 M_{\odot}$ pure baryonic NS, indicating that most of the DM is located in the cloud.

\subsubsection{Compact and Diffuse Halos}

The last column of Table \ref{tab:MRproperties} shows $M_c/M_D$, the ratio of the DM mass located in the cloud compared to the total mass of DM. We use this ratio to distinguish between two types of halos: compact and diffuse. A \textit{compact} halo has a small fraction of its DM located outside of the baryonic surface, as in the example with $y=0$ and $m_{\chi} = 0.3 \textrm{ GeV}$ with $M_c/M_D = 0.12$, and an outer radius $R_D$ that is of a similar order of magnitude as $R_B$. On the other hand, a \textit{diffuse} halo has most of its DM located in the cloud, and the outer radius is much larger than $R_B$, as is seen in the example of the case of $m_{\chi} = 0.1 \textrm{ GeV}$.

If significant amounts of DM exist in a star, i.e. $f_{\chi}$ is large, compact halos have significantly smaller $R_{B}$ compared to the pure baryonic NS with the same $M_{T} \left( R_{B} \right)$. However, if $f_{\chi}$ is small, this shortening effect on $R_{B}$ can be negligible even if $M_{c}/M_{D}$ is small, just because there is not enough DM mass to have any significant inward gravitational pull on the $R_B$. Diffuse halos, on the other hand, have $R_B$ approximately equal to that of the pure baryonic NS with
the same $M_{T} \left( R_{B} \right)$, regardless of the value of $f_{\chi}$.

\begin{figure}[ht!]
\plotone{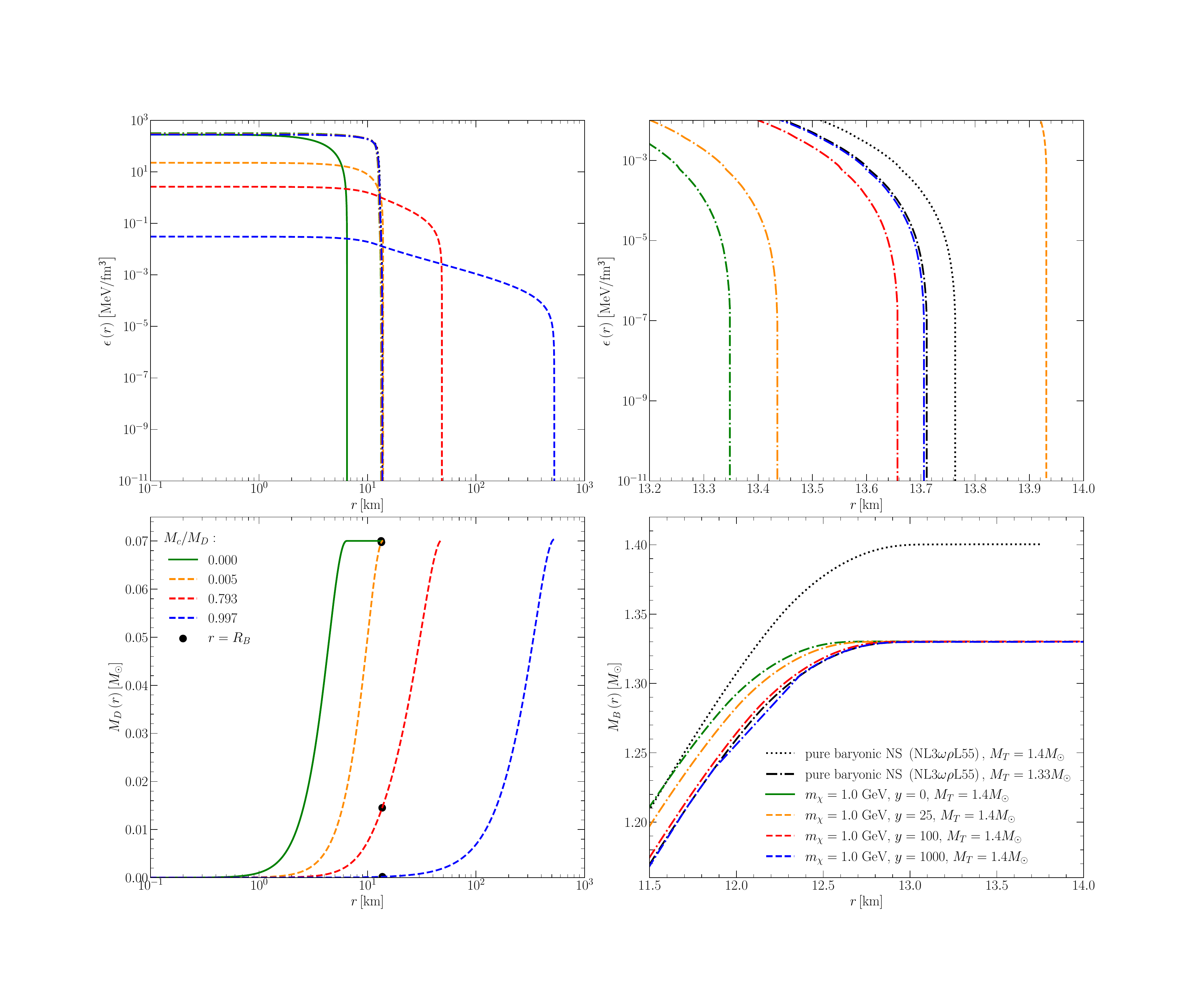}
\caption{Density and mass distributions for DANS with varying DM  self-interaction parameter $y$, and constant $m_\chi =$  1 GeV, $f_{\chi} = 0.05$, and baryonic EOS NL3$\omega \rho$L55. Descriptions of symbols and line styles are the same as for Figure \ref{fig:density-yconst}.
\label{fig:density-mconst}}
\end{figure}

Certain dark halos can have ample amounts of DM located both inside and outside their baryonic radii, in which cases a binary classification of compact or diffuse is not obvious. Such an example is the halo in Table \ref{tab:MRproperties} (and displayed in red in Figure \ref{fig:density-mconst}) resulting from $m_{\chi} = 1$ GeV and $y = 100$, with $M_{c}/M_{D} \approx 0.8$. We classify such halos as \textit{intermediate}.

Figure \ref{fig:density-mconst} shows the distribution of energy density and mass for a few DANS models with the same baryonic EOS, $M_{T} = 1.4 M_{\odot}, f_{\chi} = 0.05$  with varying $y$, and keeping the DM particle mass constant at $m_{\chi} = 1.0$ GeV, allowing for some more extreme halo examples.
Figure \ref{fig:density-mconst} (upper left and right) shows the baryonic (dash-dot) and dark (solid/dashed) energy densities. The green solid curve is the same as in Figure \ref{fig:density-yconst} representing a dark core. As repulsive self-interaction is increased between the DM particles, the pressure of the dark fluid increases, pushing the dark radius outwards to form larger and larger halos (dashed curves). Smaller repulsive self-interaction allows more DM mass to accumulate within $R_{B}$, decreasing the size of the baryonic radius. The distributions of mass in the same DANS models are shown in the lower panels. This set of DANS models includes an example of a particularly small compact halo, shown in orange ($m_{\chi} = 1 \textrm{GeV, } y = 25$), with a tiny DM cloud mass of $M_{c} = 0.005 M_{\odot}$ and a very low value of $M_{c}/R_{D} \sim 10^{-5}$. This DANS has a dark radius only half a km larger than $R_B$ (see upper right panel of Figure \ref{fig:density-mconst}).  We also feature an extreme case of a diffuse halo (shown in blue with $y=1000$) where 99.7\% of the DM mass is located in the cloud outside the baryonic surface. This case of a very diffuse halo was also shown by \citet{Miao2022}.

\section{Mass-Radius Relations} \label{sec:MR}

One of the current goals of NS astronomy is to constrain the unknown EOS of cold, dense baryonic matter through measurements of the masses and radii (or other related macroscopic quantities) of many NSs. These measurements have the potential to provide meaningful constraints, since each BM EOS is mapped to a different curve of possible NS masses and radii. A 5\% precision in the mass and radius measurements is commonly quoted \citep{Ozel2009} as a minimal requirement to provide useful constraints. The addition of dark matter converts the MR curve of a BM EOS into a two-dimensional region on the MR plane, as discussed by \citet{Kain2021}.

Many methods for estimating the NS mass and radius exist \citep{Lattimer2019}, however, in this paper we focus on the changes in methodology for gravitational self-lensing observations that are required if DM accumulates in NSs. In particular, we investigate whether the mass and radius estimates of rotation-powered ms-period X-ray pulsars made by NICER continue to be valid if the pulsars are actually DANSs instead of pure baryonic NSs. 

With the different types of masses and radii described in Section \ref{sec:model}, MR relations for DANSs can be plotted with many different combinations of the quantities, depending on the types of observations used. In this paper, we aim to connect to observations of electromagnetic radiation emitted from the baryonic surface of a DANS, so $R_B$ is most appropriate. In cases where the mass of the compact object is found through a dynamical measurement in a binary system, the total gravitational mass, $M_T$ of the DANS is the measured quantity (assuming that the size of the halo is much smaller than the orbital separation). \citet{Miao2022} suggest that in some cases, the masses determined through pulse-profile modeling correspond to $M_T(R_B)$. As we will discuss in the next section, NICER observations involve both $M_T$ and $M_T(R_B)$, so we will plot both of these masses versus $R_B$ for the different DANS EOSs.

\subsection{Stability}
\label{sec:stability}

DANS models that are stable to radial oscillations can be found using a generalized turning-point method \citep{Sorkin1982} for systems that depend on multiple parameters \citep{Jetzer1990, Henriques1990}.
Suppose an equilibrium solution to the two-fluid TOV equations is perturbed by adding or subtracting some baryon and DM particles. The change in total mass of the star is given by
\begin{equation}
\Delta M_{T} = \frac{\mu_{B}}{\sqrt{g_{tt}}} \Delta N_{B} + \frac{\mu_{D}}{\sqrt{g_{tt}}} \Delta N_{D},
\label{eq:Delta_M}
\end{equation}
where $\mu_{B}$ and $\mu_{D}$ are the BM and DM chemical potentials, respectively, and $N_{B}$ and $N_{D}$ are the number of baryon and DM particles \cite{Jetzer1990}. From the definitions of $f_\chi$ and $M_T$, it follows that
\begin{equation}
\Delta f_{\chi} = \frac{M_{B}}{M_{T}^{2}} \frac{ \mu_{D} }{\sqrt{g_{tt}}} \Delta N_{D} - \frac{M_{D}}{M_{T}^{2}} \frac{ \mu_{B} }{\sqrt{g_{tt}}} \Delta N_{B},
\end{equation}
and Equation (\ref{eq:Delta_M}) can be rewritten as
\begin{equation}
\Delta M_{T} = \frac{M_{T}}{M_{B}} \frac{\mu_{B}}{\sqrt{g_{tt}}} \Delta N_{B} + \frac{M_{T}^{2}}{M_{B}} \Delta f_{\chi}.
\end{equation}
On curves of constant DM fraction, $\Delta f_{\chi} = 0$, and at the maximum mass on these curves, $\Delta M_{T} = 0$. Therefore, $\Delta N_{B}$ is necessarily $0$ as well at the maximum mass, marking the boundary between stable and unstable equilibrium solutions. 
As a result of the turning point stability condition, curves of $M_T$ versus $R_B$ for constant values of $f_\chi$ correspond to stable DANS models if they are at, or to the right of the maximum mass on the curve. 

\begin{figure}[ht!]
\plottwo{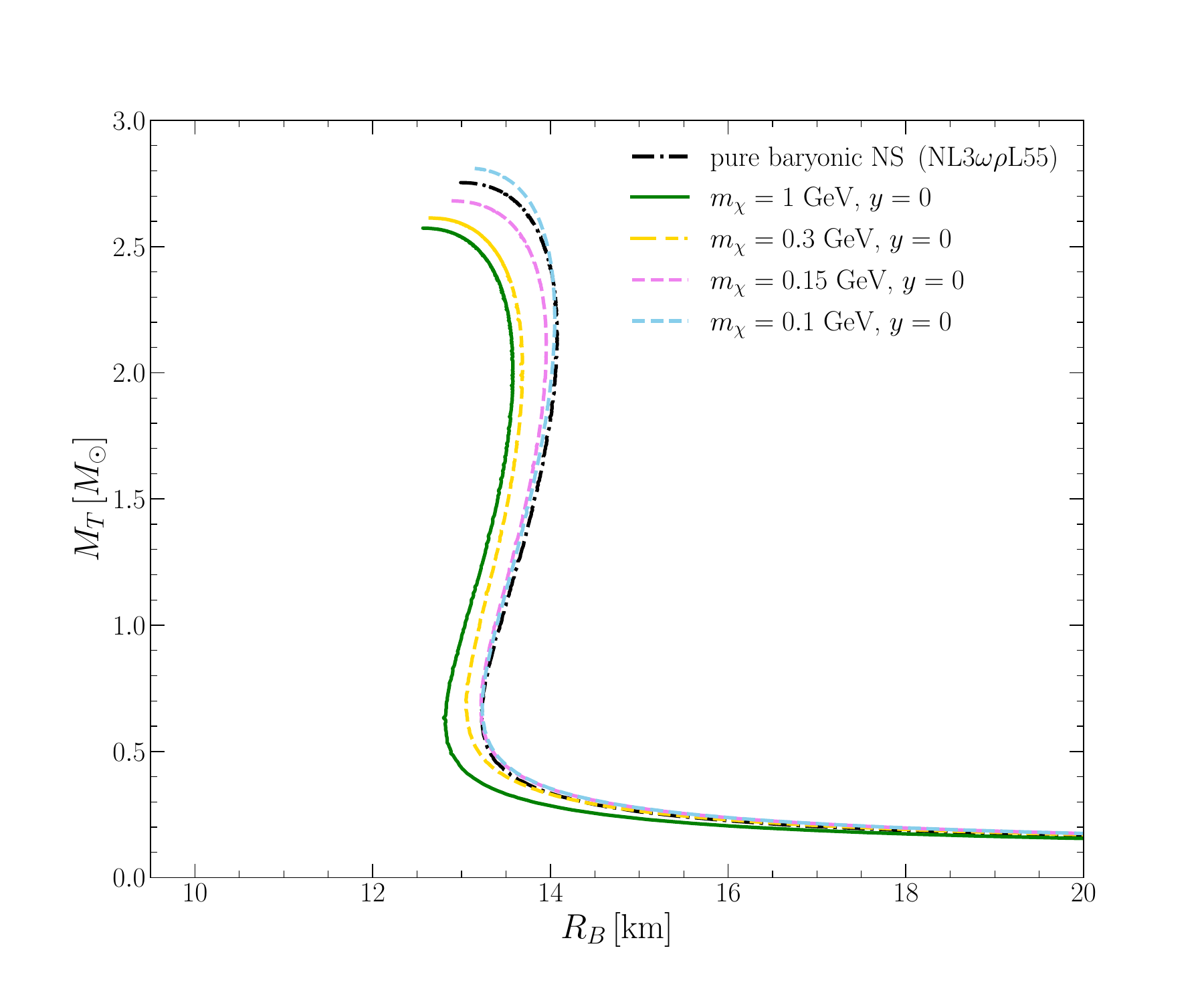}{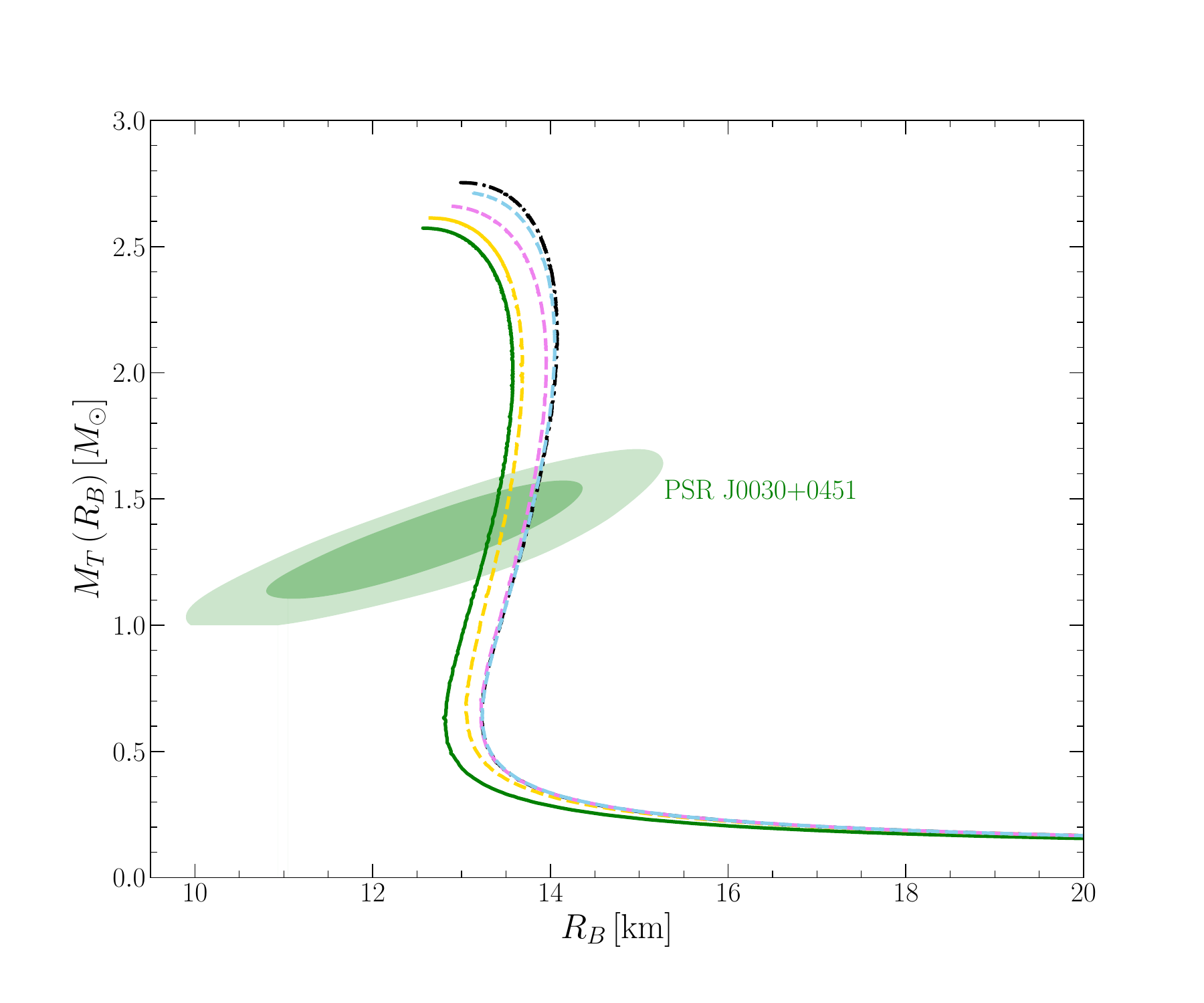}
\plottwo{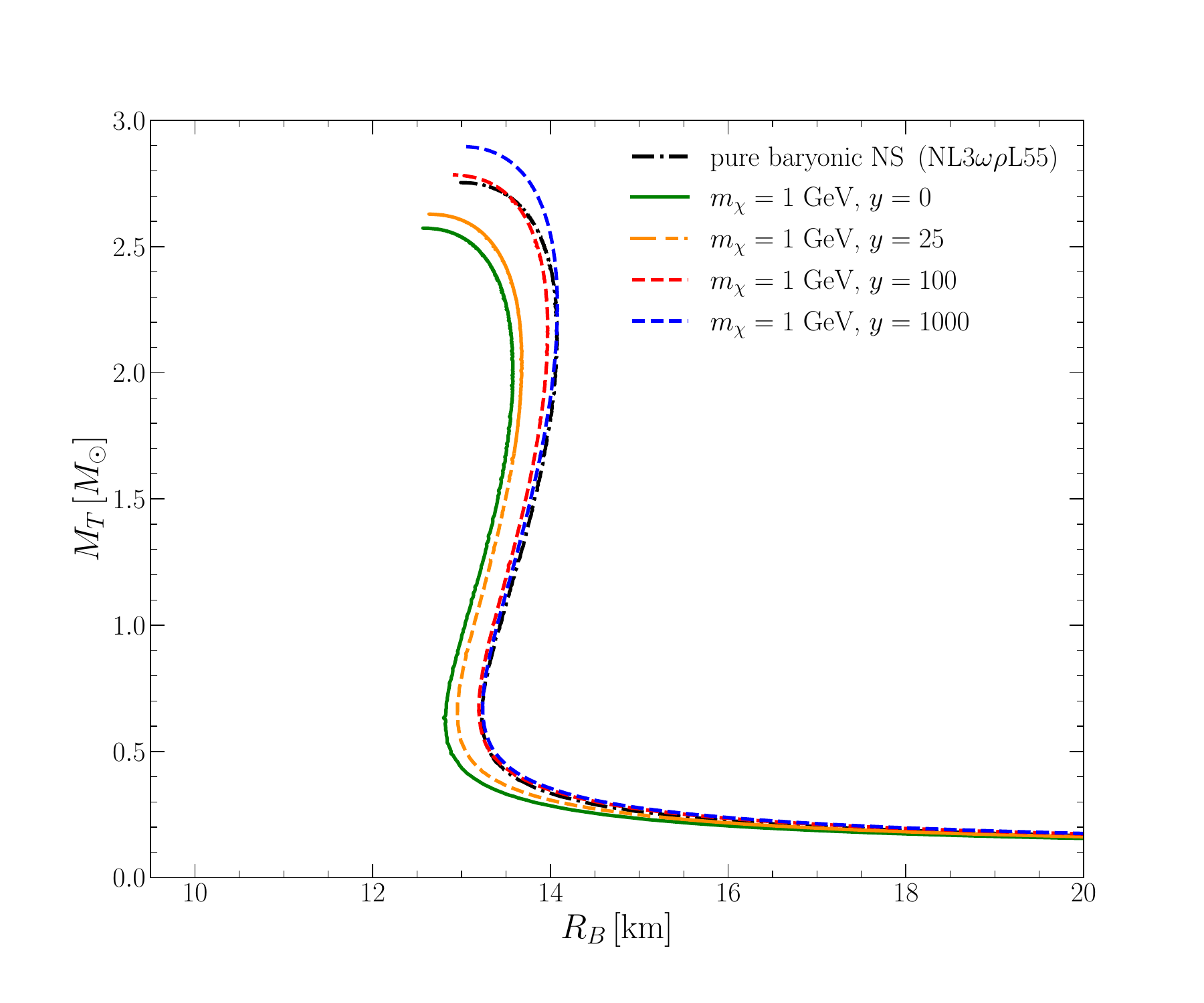}{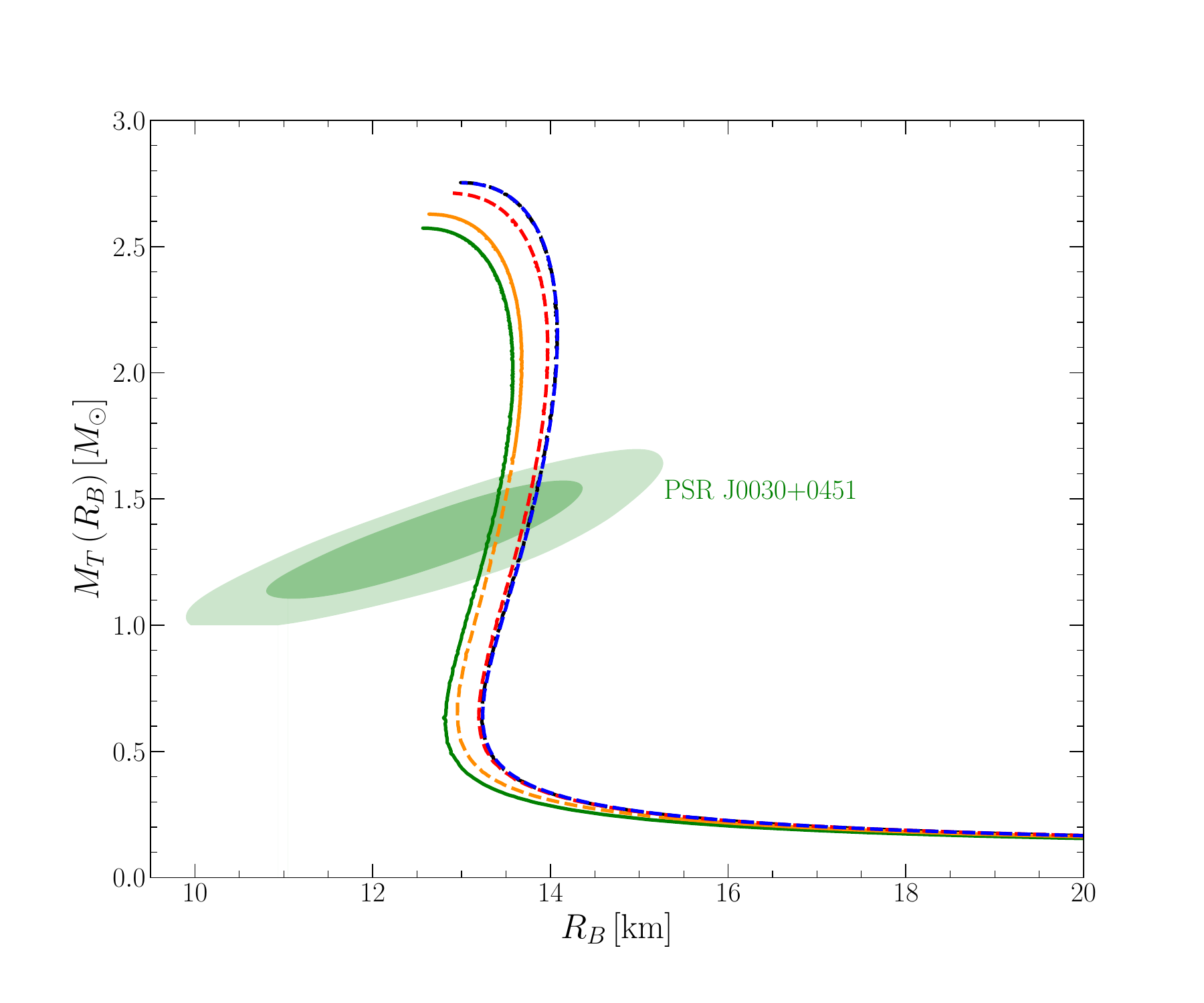}
\caption{Mass versus baryonic radius curves for baryonic EOS NL3$\omega \rho$L55 and different DM models with $f_{\chi} = 0.05$. Solid and dashed curves refer to stars with DM cores and halos, respectively.
Left: Total gravitational mass as a function of $R_B$.  Right: Total gravitational mass enclosed within $R_B$, $M_{T} \left( R_{B} \right)$, for the same DANS models. Green shaded regions show the MR constraints from one analysis of NICER data for PSR J0030+0451 \citep{Riley2019}.
\label{fig:MR-yconst}}
\end{figure}

\subsection{Types of Mass-Radius Curves}
\label{sec:mrtypes}

In Figure \ref{fig:MR-yconst} we plot either $M_T$ (left panels) or $M_T(R_B)$ (right panels) versus $R_B$ for a few sample DM particle types and the baryonic EOS NL3$\omega \rho$L55.
In the upper panels, the MR curves are for fermionic DM with 
constant $f_{\chi} = 0.05$ and $y = 0$, with varying $m_{\chi}$, while for the lower panels 
$m_{\chi} = 1$ Gev while $y$ varies.

Consider the MR curves for $M_T$ with $y=0$ shown in the upper left panel of Figure \ref{fig:MR-yconst}. For large $m_{\chi}$, dark cores tend to form and the MR curve becomes softer as a lower maximum mass is supported. As $m_{\chi}$ decreases, the curves become stiffer, supporting higher maximum masses. Stars on these stiffest curves tend to have dark halos. However, it should be noted that the curves of constant $f_{\chi}$ can have both cores and halos on them, such as the yellow curve with $m_{\chi} = 0.3$ GeV. There are also curves, such as the pink curve with $m_\chi=0.15$ GeV that correspond to DANSs with halos where the maximum mass star is smaller than the maximum allowed by the pure baryonic EOS. While we do not provide plots with varying $f_\chi$ we have verified the following behavior. For large $m_{\chi}$ and low self-interaction, increasing $f_{\chi}$ forms denser cores, softening the MR curves and allowing lower maximum masses. For large self-interaction and small $m_{\chi}$, increasing $f_{\chi}$ forms larger halos stiffening the curves. As $f_{\chi} \to 0$, the DANS curves approach the pure baryonic curve.

In Figure \ref{fig:MR-yconst} (upper right), we show $M_{T} \left( R_{B} \right)$ for the same stars. By definition, the curve representing the pure baryonic NSs (black dot-dashed curve) is the same in the left and right panels. Similarly, the curves (or parts of curves) representing dark core DANSs are the same in both panels. In the case of the yellow curve ($m_\chi = 0.3$ GeV) the lower part of the curve (dashed) corresponds to compact dark halo solutions, where a large fraction of the DM mass is inside of the baryonic surface, leading to a curve of $M_{T} \left( R_{B} \right)$ versus $R_B$ that looks very similar to the corresponding curve in the left panel. For the two DM EOS with smaller values of $m_\chi$, the $M_T(R_B)$ curves have a segment where the curve overlaps the MR curve for the pure baryonic NS. The curves where the DANS and NS curves overlap correspond to diffuse halos, where most of the DM is diffusely distributed throughout the large cloud. 
If the NSs observed by NICER are DANSs with dark halos, then NICER mass estimates are measurements of $M_{T} \left( R_{B} \right)$, not $M_{T}$ (subject to some conditions that will be discussed in Section \ref{sec:NHSC}). So, we display the $68\%$ and $95\%$ MR confidence region of the pulsar PSR J0030+0451 measured by NICER \citet{Riley2019} with the $M_{T} \left( R_{B} \right)$ curves, but not the $M_{T}$ curves. The case of PSR J0740+6620 (which 
 will be discussed in Section \ref{sec:bothNICERradio}) is more complicated, so we do not show this pulsar's NICER MR confidence regions on either diagram.

Similar $M_T$ and $M_T(R_B)$ versus $R_B$ curves are shown in the lower panels of Figure \ref{fig:MR-yconst} for fixed values of $m_\chi$ and different values of the self-interaction parameter $y$. Increasing $y$ increases the size of the halo and the maximum value of $M_T$, but not $M_T(R_B)$. In the case of $y=1000$ (blue dashed curve), the $M_T(R_B)$ curve is virtually coincident with the pure baryonic MR curve (black dot-dashed curve), so they are indistinguishable on this plot. For the other curves, significant amounts of DM exist within the baryonic radii of the stars. This causes a decrease of $R_{B}$ and a lower mass enclosed within it. We find no $M_{T} \left( R_{B} \right)$ curves which are stiffer than the pure baryonic MR curve, for any value of $f_\chi$. This is because any significant DM mass within $R_{B}$ will always decrease $R_{B}$, resulting in lower $M_{T} \left( R_{B} \right)$ due to the more tightly-bound configuration. Thus, the pure baryonic MR curve provides an approximate upper boundary for $M_{T} \left( R_{B} \right)$ curves for DANSs. These properties of diffuse halos do not depend on the properties of the dark matter in the following sense. If a chosen dark matter model (which could be bosonic) results in a DANS with a diffuse halo and a DM fraction $f_\chi$, the maximum total mass increases by the same amount and the $M_T(R_B)$ curve will coincide with the pure baryonic MR curve.

Generally, if the maximum mass DANS on a constant $f_\chi$ MR curve has a dark halo, all of the DANS with lower mass on the same curve will also have halos. Similarly, if the maximum mass DANS has a diffuse halo, the lower mass DANS will also have diffuse halos. However, if the maximum mass DANS has a dark core, it is possible to find a transition to halo solutions for smaller masses. Additionally, if the maximum mass DANS has a compact halo, a transition to a diffuse halo at some lower mass is possible. Constant $f_{\chi}$ curves where this transition happens, are on or near the boundary of regions of cores and halos.

These properties of a DANS MR curve for fixed $f_\chi$ can be understood by considering the properties of the DANS with the maximum value of $M_T$ on the curve. The reformulation of the TOV equations using enthalpy as the independent variable introduced by \citet{Lindblom1992}, allows a simple test to determine whether a DANS is a core or halo model. \citet{Miao2022} showed that comparing the enthalpies of the two non-interacting fluids at the centre of the DANS leads to the simple criterion that the fluid with the larger central enthalpy will have the larger radius. As a result if the DM has a larger central enthalpy, then a dark halo forms. If one considers the pure baryonic MR curve, the central baryonic enthalpy decreases as mass decreases on the curve. Adding DM to a NS makes only a very small change to the baryonic enthalpy, so it continues to be true that the largest baryonic central enthalpy coincides with the maximum value of $M_T$ on a curve of constant $f_\chi$. 
When the maximum mass DANS has a dark core, then this star has a DM central enthalpy smaller than the baryonic central enthalpy. 
However, further down the MR curve of constant $f_\chi$, the baryonic enthalpy will be decreasing making it possible for the curve to transition to halo models for lower values of mass. 
If the maximum mass DANS has a dark halo, then the DM central enthalpy is larger than the central baryonic enthalpy. The baryonic central enthalpy decreases down the MR curve. If the dark enthalpy does not decrease as rapidly, the DANS with lower masses continue to have dark halos.
We find that if the maximum mass star has a diffuse halo, its baryonic central enthalpy falls more rapidly than the dark enthalpy as one moves towards lower mass stars (on a curve of constant $f_\chi$). 
So if the maximum mass DANS has a diffuse halo, the lower-mass DANS on the same MR curve also have diffuse halos. 
For curves near the boundary of cores and halos, it is possible for the central dark enthalpies to decrease faster than the central baryonic enthalpies resulting in a transition from halos to cores on a constant $f_{\chi}$ curve ($M_{T} \lesssim 0.3 M_{\odot}$ of the orange curves in Figures \ref{fig:MR-yconst} and \ref{fig:BM-EOSeffects} (left)).

\subsection{Effect of the Baryonic EOS}
\label{sec:baryoniceos}

The maximum mass NS has a central baryonic enthalpy that does not vary much for different baryonic EOSs. For example, the central enthalpies for the maximum mass NS for the two baryonic EOSs used in this paper differ by 3\%. In the case of a diffuse halo, the central dark enthalpy is much larger than the baryonic enthalpy for any baryonic EOS. This is equivalent to the statement that the radii of baryonic NS only vary by a couple of km and are all of order 10 km. Meanwhile, $R_D$ is typically quite large ($\sim$ 100 kms) compared to $R_B$ if the halo is diffuse. As a result, the different baryonic radii for different baryonic EOSs do not significantly affect the properties of a dark halo if it is diffuse.

\begin{figure}[ht!]
\plottwo{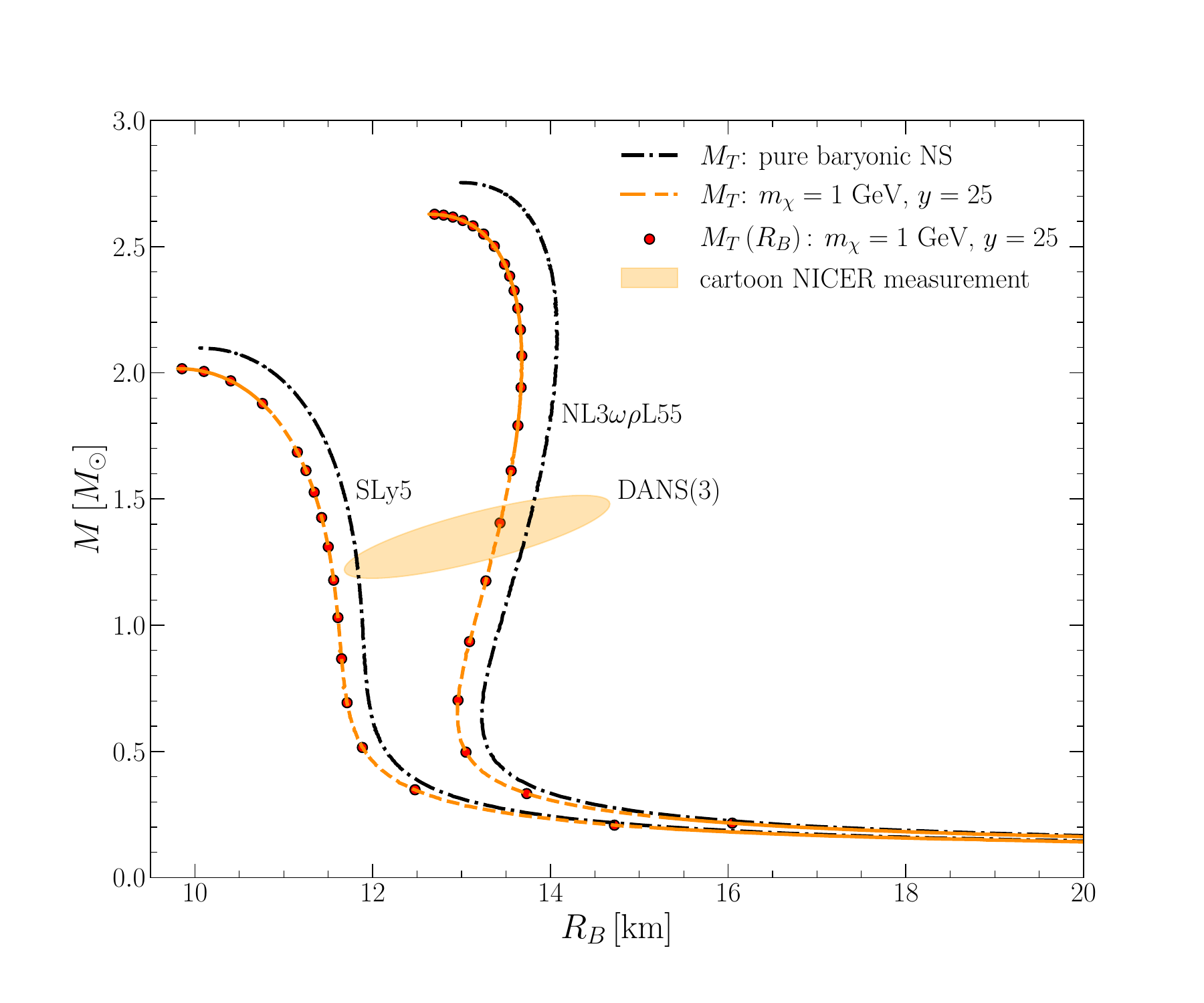}{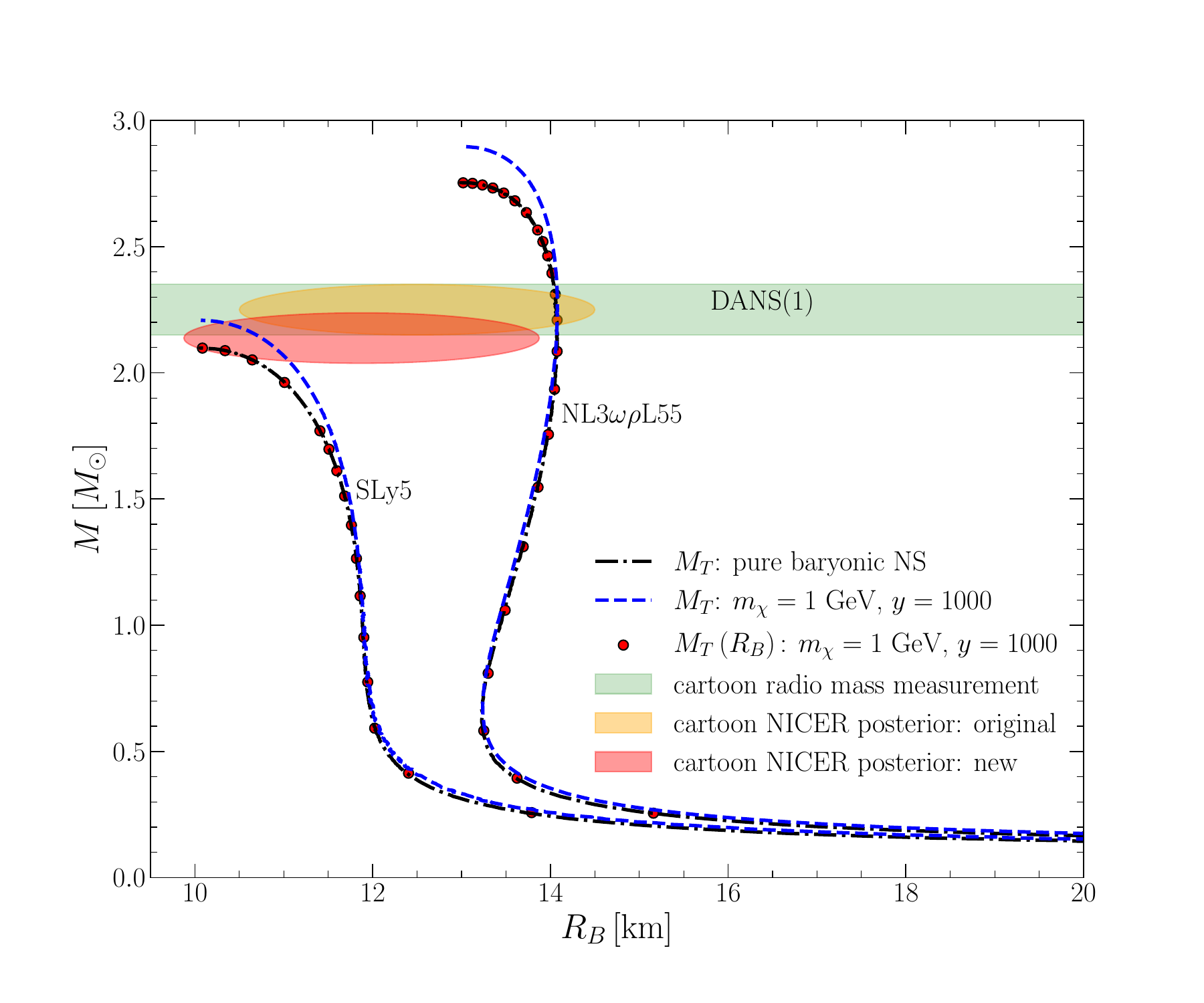}
\caption{Effects of different BM-EOSs on the $M_T(R_B)$ curves of compact (left) and diffuse (right) halos with $f_{\chi} = 0.05$. Cartoon MR constraints for two fictitious DANSs are shown for illustrative purposes. Left: For both BM-EOSs, the curve of $M_T(R_B)$ versus $R_B$ (red dots) is almost identical to the $M_{T}$ versus $R_{B}$ curve for the DANSs (orange curves), due to the tiny amounts of DM in the cloud. The orange-shaded region depicts a cartoon  1$\sigma$ confidence region 
for a gravitational self-lensing observation of a fictitious DANS by a telescope such as NICER.
 For the BM-EOS NL3$\omega \rho$L55, both pure baryonic EOS and DM-EOS are supported by the observations of DANS(3). The pure BM-EOS SLy5 is suported by the observations, but the corresponding DM-EOS is ruled out. Right: For both BM-EOSs, the curve of $M_T(R_B)$ versus $R_B$ (red dots) is almost identical to the MR curve for the purely baryonic NS with no DM content (solid black curves). The green shaded region depicts a cartoon 1$\sigma$ confidence region for a dynamical NS mass measurement of a different DANS, DANS(1). 
 The orange-shaded region depicts a cartoon  1$\sigma$ gravitational-lensing MR posterior for DANS(1), obtained using a mass prior using the dynamical mass measurement. The red-shaded region depicts an example of a different 1$\sigma$ posterior obtained using an adjusted mass prior that tests for nonzero values of $f_{\chi}$. For more details on the interpretation of this DANS, see Section \ref{sec:bothNICERradio}.
\label{fig:BM-EOSeffects}
}
\end{figure}

In Figure \ref{fig:BM-EOSeffects} we show the effect of changing the baryonic EOS while keeping the DM EOS constant. In Figure \ref{fig:BM-EOSeffects} (left), a DM EOS leading to compact halos for both the stiff EOS NL3$\omega \rho$L55  and the soft EOS SLy5 is shown for $f_\chi=0.05$. Tiny amounts of DM exist in the clouds of the DANSs ($M_{c}/M_{D} < 0.025$ for $M_{T} \geq 1 M_{\odot}$), and the red $M_{T} \left( R_{B} \right)$ dots line up with the orange $M_{T}$ curves. 
Increasing $f_{\chi}$ for these compact halos will result in them turning into dark cores (stars near the maximum masses of these curves are already dark cores). The curves presented here are examples of the transition region between dark cores and compact dark halos.

In Figure \ref{fig:BM-EOSeffects} (right), a DM EOS leading to a diffuse halo for all masses for the stiff EOS 
NL3$\omega \rho$L55 is shown for $f_\chi=0.05$. The MR curve for a much softer EOS, SLy5, is shown using the same DM EOS and DM fraction. For both baryonic EOSs, the pure baryonic MR curves are shown as solid black curves, while the $M_T$ versus $R_B$ curves are shown as blue dashed curves. The corresponding $M_T(R_B)$ curves are displayed using red dots so that it can be seen that these curves overlap with the pure NS MR curves for both baryonic EOS.

\begin{figure}[ht!]
\plottwo{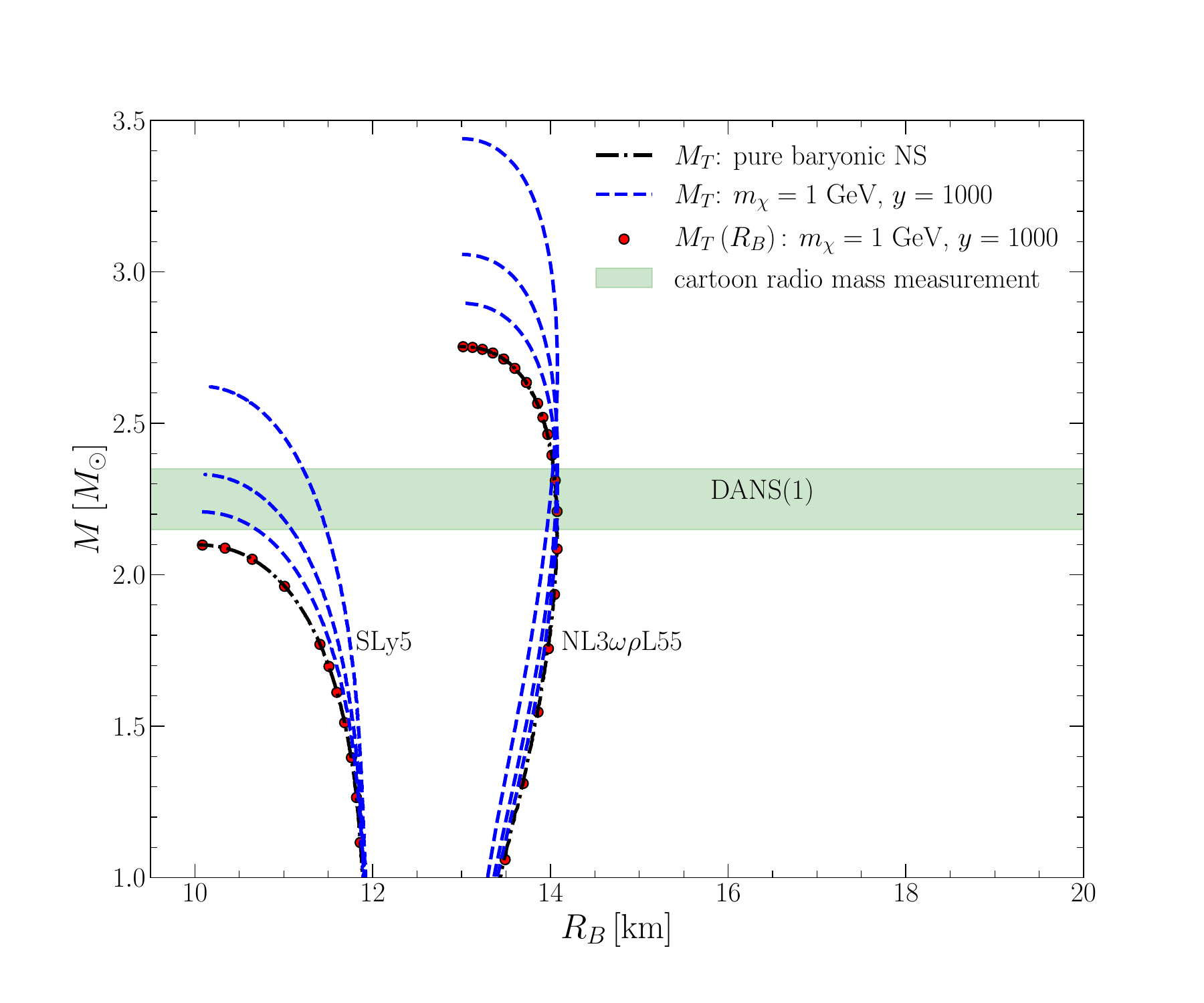}{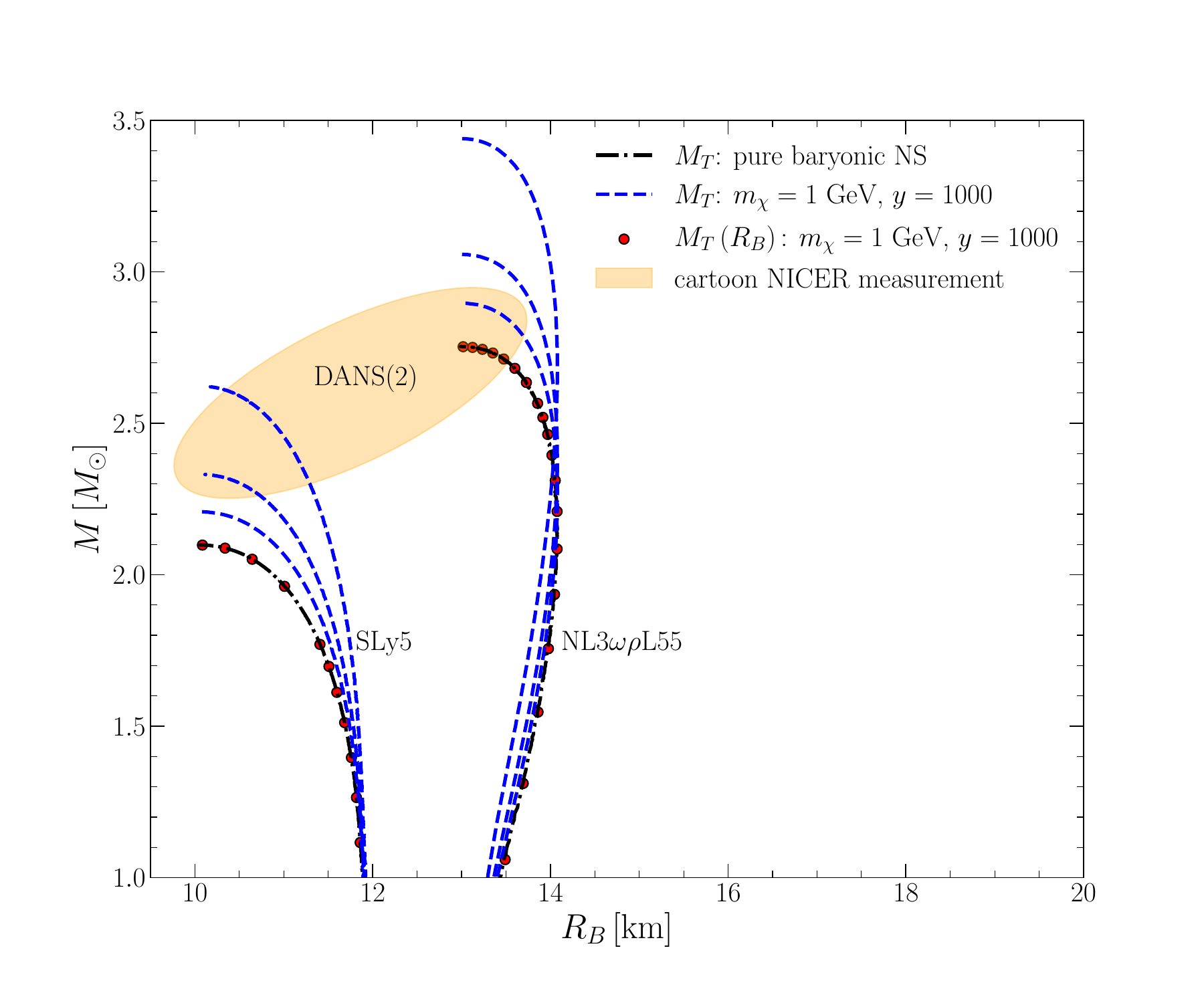}
\caption{Effect of increasing dark matter fraction $f_\chi$ on MR curves for diffuse halos. These schematic diagrams show how to interpret the validity of diffuse halo MR curves when only a dynamical mass measurement of NS (left) or only a gravitational self-lensing MR constraint is available (right). The blue dashed curves correspond to $M_T$ for dark matter mass fractions $f_\chi = 0.05, 0.1, 0.2$ (where higher mass corresponds to larger values of $f_\chi$). 
Left panel: 
The green shaded region depicts a cartoon 1$\sigma$ confidence region of dynamical mass measurement of DANS(1). The dynamical mass measurement supports a BM-EOS if a $M_T$ curve (blue dashed lines) intersects the range of measured masses. The dynamical mass measurement of DANS(1) supports the EOS NL3$\omega \rho$L55 for any value of $f_\chi$ while EOS SLy5 is only supported if the value of $f_\chi$ is large enough.
Right panel: The orange shaded region depicts a cartoon 1$\sigma$ NICER confidence region for a different NS MR measurement. The observation supports EOS NL3$\omega \rho$L55 with any value of $f_\chi$ since the $M_T(R_B)$ vs $R_B$ curves (red dots) for varying $f_\chi$ are identical for diffuse halos. The observation rules out EOS SLy5 for any diffuse halo and any value of $f_\chi$.
}
\label{fig:only_radio_or_NICERdiffuse}
\end{figure}

Figure \ref{fig:only_radio_or_NICERdiffuse} shows the effect of changing $f_\chi$ on a diffuse halo, using the same EOSs.
The blue dashed curves of $M_T$ versus $R_B$ change, with maximum values of $M_T$ that increase as $f_\chi$ increases. In particular, since almost all of the dark matter is located in the cloud, the difference in masses for the maximum mass star is $M_{T,\text{max}} - M_{T,\text{max}}(R_B) \simeq M_D$. If the maximum mass pure NS is denoted $M_{\text{TOV}}$, then adding a diffuse dark matter halo to create a DANS with mass fraction $f_\chi$ increases the maximum total mass by $\Delta M/M_{\text{TOV}} = (M_{T,\text{max}} - M_{\text{TOV}})/M_{\text{TOV}} = M_D/M_B = f_\chi/(1-f_\chi)$. For example, in Figure \ref{fig:only_radio_or_NICERdiffuse} for both of the two different baryonic EOSs shown, it can be seen that adding a diffuse dark matter halo with $f_\chi = 0.05$ has increased the maximum total mass by slightly more than 5\%  (the actual increases are 5.2\%). However, for all these constant $f_\chi$ curves, the $M_T(R_B)$ versus $R_B$ curves all overlap and are independent of $f_\chi$. The feature that these curves are independent of $f_\chi$ for diffuse halos is an important property for the analysis proposed in Section \ref{sec:methods}.

\section{Gravitational Self-Lensing for a DANS with a Dark Matter Halo} \label{sec:lens}

Due to the strong gravitational potential near $R_{B}$, NSs gravitationally lens the light they emit from their baryonic surface. X-ray timing missions like NICER take advantage of this self-lensing property of NSs to infer their masses and radii \citep{Watts2016}. In the case of a DANS with a dark halo, the X-rays emitted from $R_{B}$ may experience extra bending due to the DM distributed outside $R_{B}$. The extra light-bending depends on the amount of mass in the cloud and the distribution of mass within the cloud.

We work in the Schwarzschild plus Doppler (SD) approximation \citep{Miller1998, Poutanen2003} in this paper to analyse the self-lensing properties of DANSs. In the SD approximation, the surface of the star is approximated by a sphere, and the spacetime metric outside of the star's surface is approximated by the Schwarzschild solution with the same mass.  For stars that rotate with frequencies between 200 - 400 Hz (as is true for the pulsars observed by NICER), adding the oblate shape of the star through the Oblate Schwarzschild (OS) approximation \citep{Morsink2007} provides a sufficiently accurate method for computing the gravitationally-lensed emission. The OS approximation relies on a quasi-universal parametrization of the oblate shape that was derived for purely baryonic NSs. 
This parametrization holds for DANSs with a DM core \citep{2024arXiv240501487Konstantinou}, and 
we conjecture (but have not yet proved) that this parametrization will also apply to DANS with a halo. If the oblate shape parameterization is not changed significantly by the DM, then the extension of these results to an oblate DANS will be straightforward. We will address this question elsewhere and instead focus in this paper on the changes to gravitational lensing by a spherical star to understand the relative importance of the introduction of DM. 

\subsection{Light-bending Geometry} \label{sec:SD}

The geometry for gravitational lensing is shown in Figure 1 of \citet{Bogdanov2019b}, and we use the same notation in this paper. We will only consider an infinitesimal hot spot since larger, more complicated spots can be built up by the addition of small spots. The star's spin frequency, as measured by an observer at infinity is $\nu$, and the observer's co-latitude, relative to the star's spin axis is $\zeta$. The angle $\zeta$ is normally assumed to coincide with the usual inclination angle definition for binary systems, although this restriction can be relaxed \citep{Miller2021}. The co-latitude of the spot's centre is $\theta_c$, and its longitude $\phi(t) = 2 \pi \nu t$, where $\phi=0$ is defined by the longitude of the observer. Given these definitions, the bending angle, $\psi$ at any moment is given by the geometric relation
\begin{equation}
\cos{\left( \psi(t) \right)} = \cos{\left( \zeta \right)} \cos{\left( \theta_c \right)} + \sin{\left( \zeta \right)} \sin{\left( \theta_c \right)} \cos{\left( \phi(t) \right)}.
\end{equation}

The bending angle $\psi$ is related to the photon's emission angle, $\alpha$ relative to the normal to the surface by the integral from the baryonic surface to infinity,
\begin{equation}
\psi(\alpha) = (1+z_B) \sin{\left( \alpha \right)} \int_{0}^{1} du \;
e^{\Phi(u)+\Lambda(u)}
 \bigg \{
 1 - \left[ \left( 1 + z_B \right) \sin{\left( \alpha \right)} \; u \; e^{\Phi(u)} \right]^2
  \bigg \}^{-1/2},
  \label{eq:psi}
\end{equation}
where $u=R_B/r$, $z_B$ is the gravitational redshift from the baryonic surface, and $\Phi$ and $\Lambda$ are the metric potentials defined in Equation (\ref{eq:metric}). Equation (\ref{eq:psi}) for the bending angle agrees with the equation derived by \citet{Miao2022} (although we use different notation) and agrees with the standard Schwarzschild result \citep{Pechenick1983} when there is no halo. Note that the coordinate $u$ has the value of 1 at the baryonic surface, 0 at infinity, and $R_B/R_D$ at the outer edge of the DM halo.
Equation (\ref{eq:psi}) is inverted to solve for the angle $\alpha$ given a bending angle $\psi$. The divergence of light rays,
$d\psi/d\alpha$, given by
\begin{equation}
    \frac{d\psi}{d\alpha} = (1+z_B) \cos{\left( \alpha \right)} \int_{0}^{1} du \;
e^{\Phi(u)+\Lambda(u)}
 \bigg \{ 
  1 - \left[ \left( 1 + z_B \right) \sin{\left( \alpha \right)} \; u \; e^{\Phi(u)} \right]^2
  \bigg \}^{-3/2}
  \label{eq:dpsidalpha}
\end{equation}
is similar in form to the corresponding equation for a NS without dark matter.

The flux of photons, $dF(E)$ with observed energy $E$ and emitted at angle $\alpha$ from the normal to the surface can be generalized from \citet{Bogdanov2019b} to the case of a DANS as
\begin{equation}
    dF(E) = \frac{\delta^4}{(1+z_B)} I'(E',\alpha') \cos{\left( \alpha \right)} \frac{d \cos{\left( \alpha \right)}}{d \cos{\left( \psi \right)}} \frac{dS'}{D^2}, 
    \label{eq:dFE}
\end{equation}
where primes are used to denote quantities in the frame co-moving with the star, $I'$ is the emitted specific intensity, and $dS'/D^2$ is the ratio of the surface area of the spot in the co-moving frame (see \citet{Bogdanov2019b} for more details on $dS'$) to the square of the distance to the observer and does not depend on time. The Doppler factor, denoted by $\delta$ depends on time through the equation
\begin{equation}
\delta = \frac{1}{\gamma \left[ 1 - v \cos{\left( \xi \right)} \right]},
\end{equation}
where
\begin{equation}
v = {2 \pi \nu R_{B} \sin{\left( \theta_c \right)}} (1 + z_B)
\label{eq:v}
\end{equation}
is the local speed of the baryonic surface with $\gamma = (\sqrt{1 - v^{2}})^{-1/2}$, the Lorentz factor, and
the angle between the velocity vector and the emitted photon, $\xi$, is defined by
\begin{equation}
\cos{\left( \xi \right)} = \frac{\sin{\left( \alpha \right) \sin{\left( \zeta \right)} \sin{\left( \phi \right)}}}{\sin{\left( \psi \right)}}.
\end{equation}
The photon energy in the observer's frame, $E$, is related to the energy $E'$ in the frame co-moving with the star by $E = \delta (1+z_B)^{-1} E'$, and the photon emission angles are related by $\cos{\left( \alpha' \right)} =\delta \cos{\left( \alpha \right)}$. The Doppler boost factor is affected by a dark matter halo through the change in the gravitational redshift factor entering in the definition of the velocity in Equation (\ref{eq:v}) and the change in the angle $\xi$ introduced by changes in the relation $\psi(\alpha)$. 

The final ingredient left to plot pulse profiles is the observer time defined by
\begin{equation}
t_{\text{obs}} - t = \Delta t  = R_B \int_{0}^{1}  \frac{du}{u^2} \; e^{\Lambda(u) - \Phi(u)}
 \left( \bigg \{ 1 - \left[ \left( 1 + z_B \right) \sin{\left( \alpha\right)} \; u \; e^{\Phi(u)} \right]^2 \bigg \}^{-1/2} - 1 \right).
 \label{eq:deltat}
\end{equation}

\subsection{Bolometric Light Curves}
\label{sec:bolo}

The parameter estimation analysis performed to determine the masses and radii of X-ray pulsars using NICER data requires that photon-energy-dependent waveforms be calculated to disentangle the effects of light-bending and atmospheric beaming. However, the effect of dark matter on the pulse profile does not depend on photon energy, so for the purpose of evaluating the magnitude of the changes caused by dark matter, it is sufficient to consider bolometric light curves for isotropic emission from the baryonic surface. We will consider the effect of dark matter on realistic atmosphere models in Section \ref{sec:atmos}.

The normalized bolometric flux, $F$, results after integrating the energy-depended flux in Equation (\ref{eq:dFE}) over the photon energy in the observer's frame and dividing by the phase-independent terms that are unaffected by a dark matter halo,
\begin{equation}
F = \frac{\delta^5}{(1+z_B)^2} \cos{\left( \alpha \right)} \frac{d \cos{\left( \alpha \right)}}{d \cos{\left( \psi \right)}}.
\label{eq:F}
\end{equation}
The most important difference between the bolometric and energy-dependent fluxes is the extra factor of the gravitational redshift $(1+z_B)$ that is added to the bolometric flux through the integration over observed photon energy, since the gravitational redshift changes the overall normalization of the flux.

\begin{figure}[ht!]
\plottwo{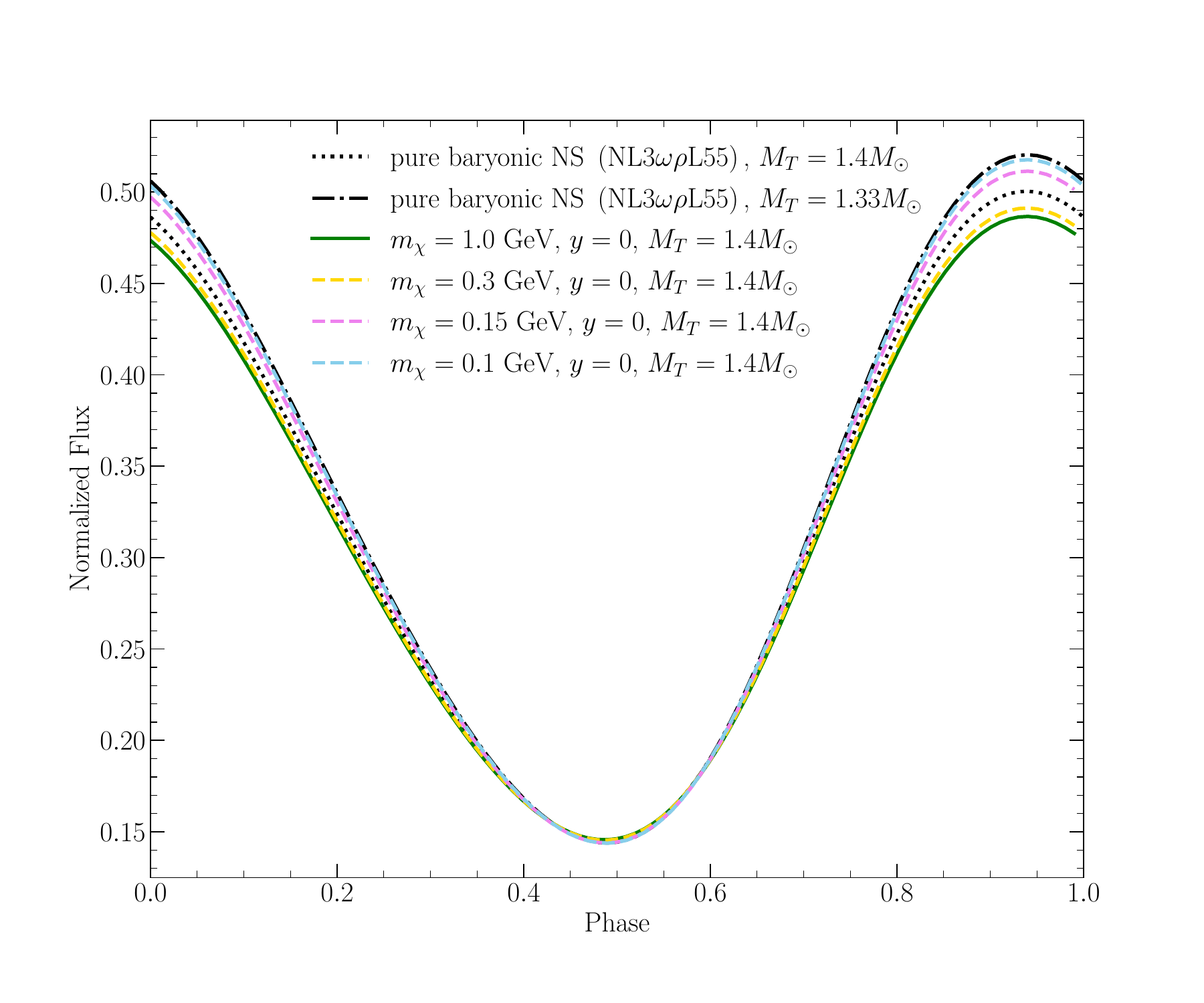}{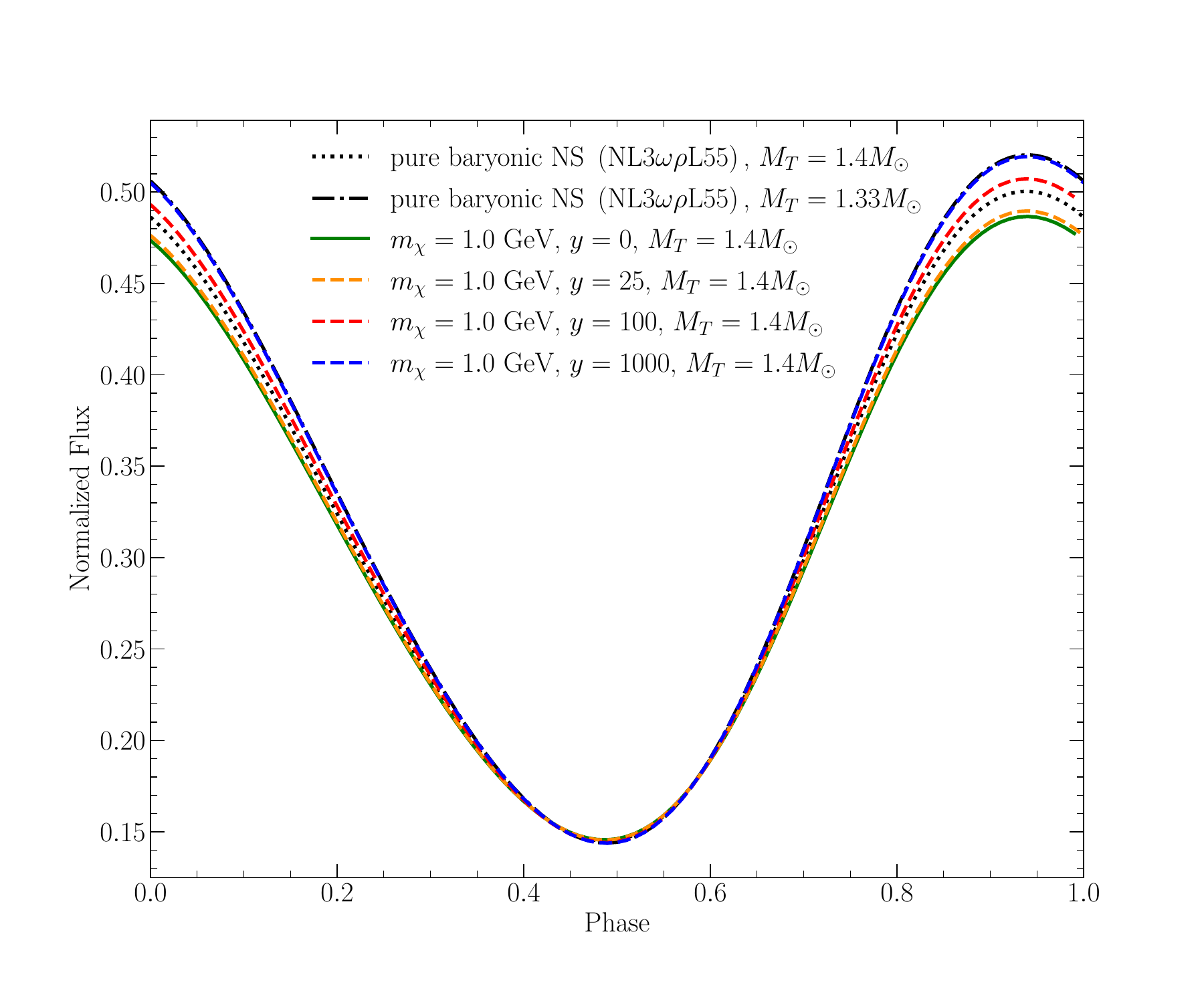}
\caption{Pulse profiles for the stars in Figures \ref{fig:density-yconst} (left) and \ref{fig:density-mconst} (right) with total mass $M_T = 1.4 M_\odot$. All stars have a rotation frequency of $\nu = 200$ Hz in the observer's frame and $\theta_c = \zeta = \pi/4$.
\label{fig:F_vs_t}}
\end{figure}

Figure \ref{fig:F_vs_t} shows the bolometric flux as a function of observed pulsar phase ($t_{obs} \times \nu)$ produced by the same stars shown in Figures \ref{fig:density-yconst} (left) and \ref{fig:density-mconst} (right). The DANSs all have a total mass of 1.4~$M_\odot$, rotate at frequency of $\nu = 200$ Hz and have an infinitesimal hot spot located at the northern hemisphere colatitude of $\theta_c = \pi/4$. The inclination angle of the spin axis relative to the line of sight is chosen to be $\zeta = \pi/4$ for all the stars.
Choosing $\zeta = \theta_c$ allows the light curve to sample photons emitted from a wide range of angles, $\alpha$, ranging from $\alpha=0$ (normal to the surface) to smaller angles corresponding to emission from close to the star's limb. Changes to the rotational frequency have the effect of shifting the phase of the maximum flux by a small amount.

The maximum observed flux from each of these stars is directly proportional to the value of the time-time component of their metric at $R_{B}$, $g_{tt} \left( R_{B} \right) = e^{2\Phi_B} = (1+z_B)^{-2}$. 
For each set of pulse profiles shown in Figure \ref{fig:F_vs_t}, the profile with the lowest maximum flux is for a DANS with a core, and as the maximum flux increases, the curves correspond to DANSs with halos that are increasingly more diffuse. As the halos become more diffuse, the light curves approach the light curve for a pure baryonic NS  with the same $M_{B}$ and $R_{B}$ as the DANS. This is because the more diffuse the halo becomes, the closer its metric approaches the Schwarzschild metric. Since $f_{\chi} = 0.05$ for these $M_{T} = 1.4 M_{\odot}$ DANSs, the flux from the $M_{T} = 1.33 M_{\odot}$ star forms the upper boundary of the flux observed from diffuse halos. Stars with sufficiently diffuse halos closely mimic the flux observed from stars with the same $R_{B}$ and $M_{T} \left( R_{B} \right)$ but no DM clouds (i.e. spacetime is exactly Schwarzschild outside $R_{B}$). We call such stars with no DM cloud, the no-halo Schwarzschild counterpart to the DANS. Thus, to find an appropriate measure of change in observed flux introduced by the DM cloud for both compact and diffuse halos, we need to compare DANSs with their no-halo Schwarzschild counterparts as was first proposed by \cite{Miao2022}. Note that these \textit{no-halo} stars are not necessarily solutions resulting from any particular BM-EOS. They are just a reference star with the same $M_{T} \left( R_{B} \right)$ and $R_{B}$ as their DANS counterpart, but with no DM cloud (their total mass is $M_{T} = M_{T} \left( R_{B} \right)$).

\begin{figure}[ht!]
\plottwo{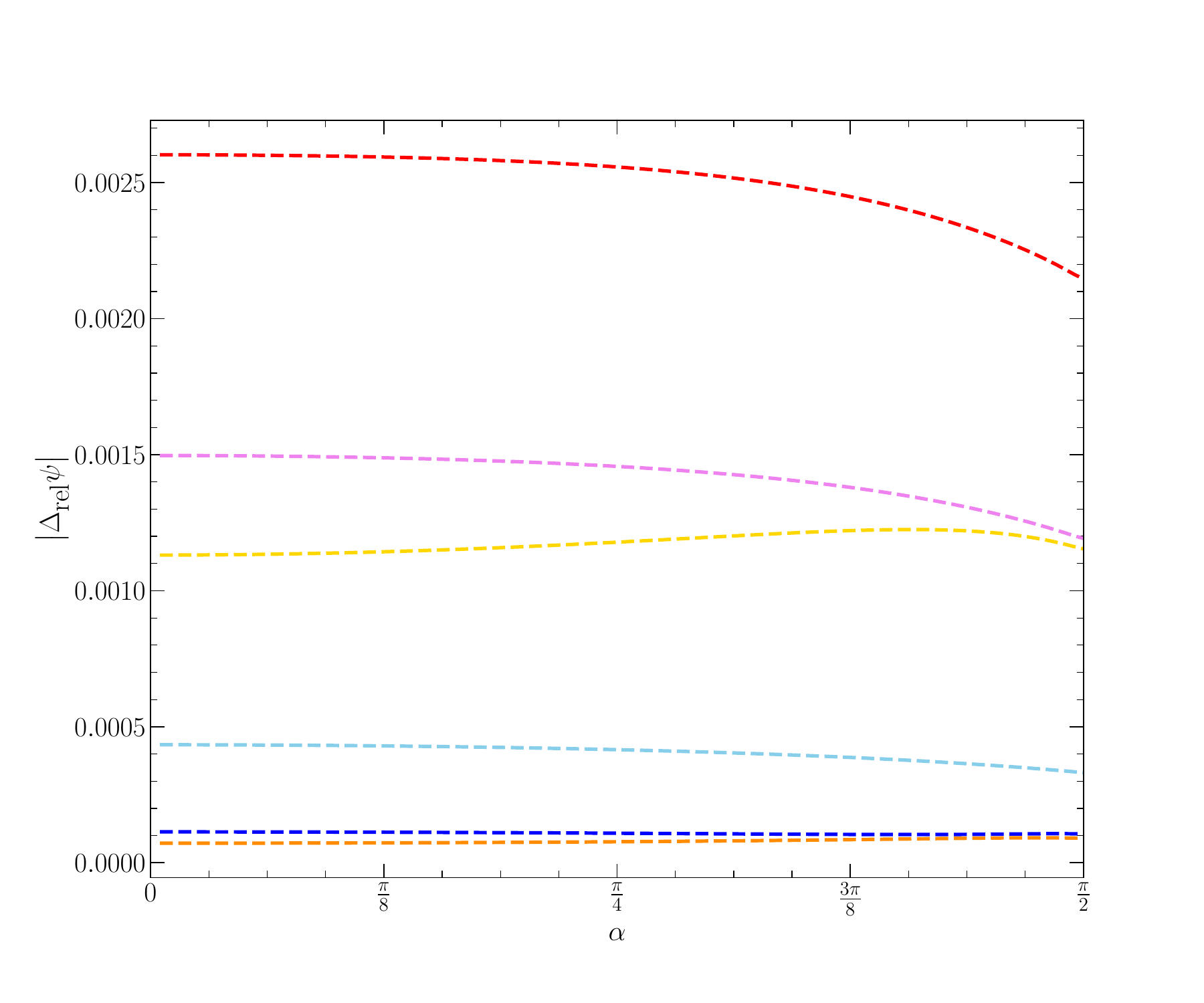}{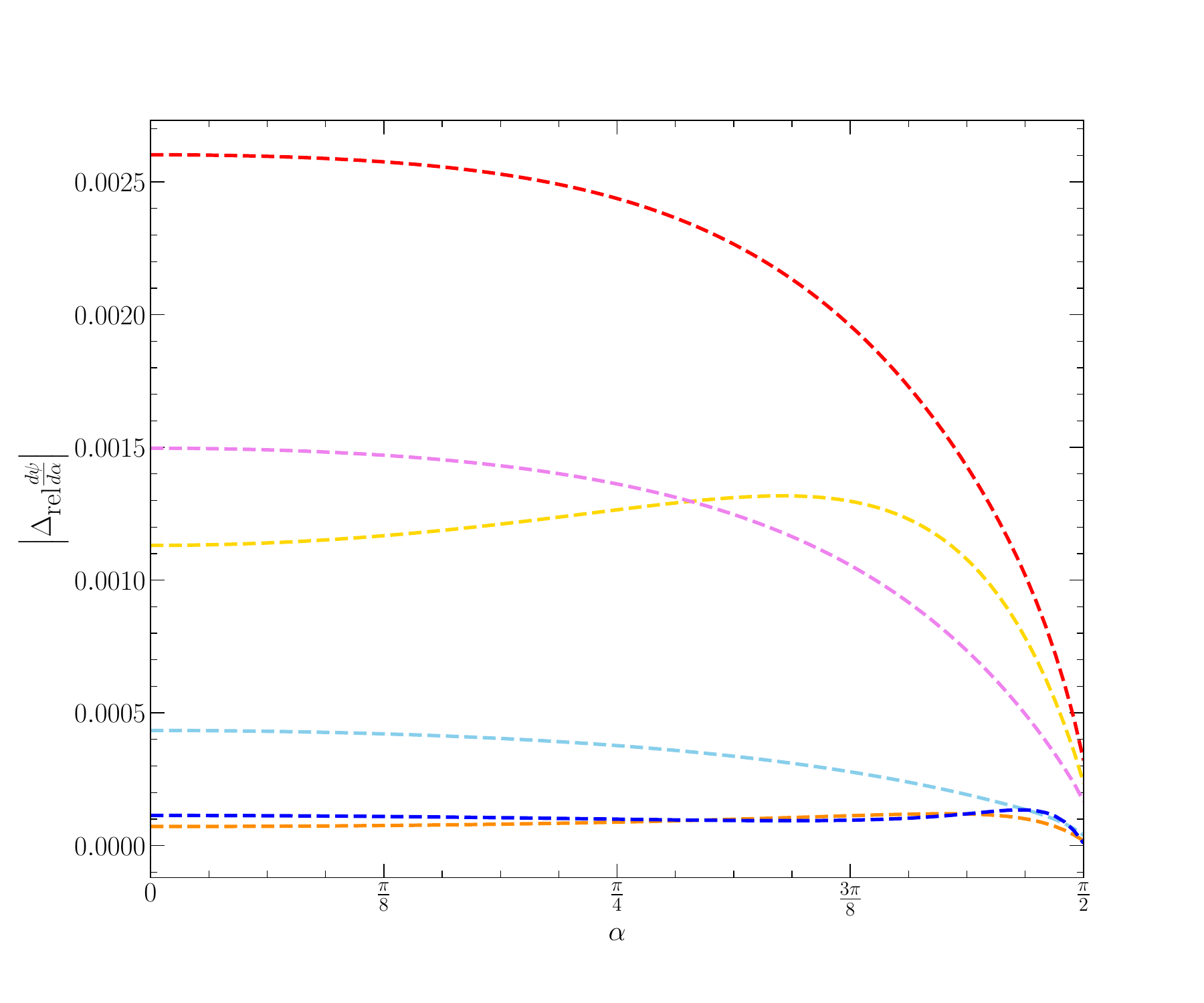} \plottwo{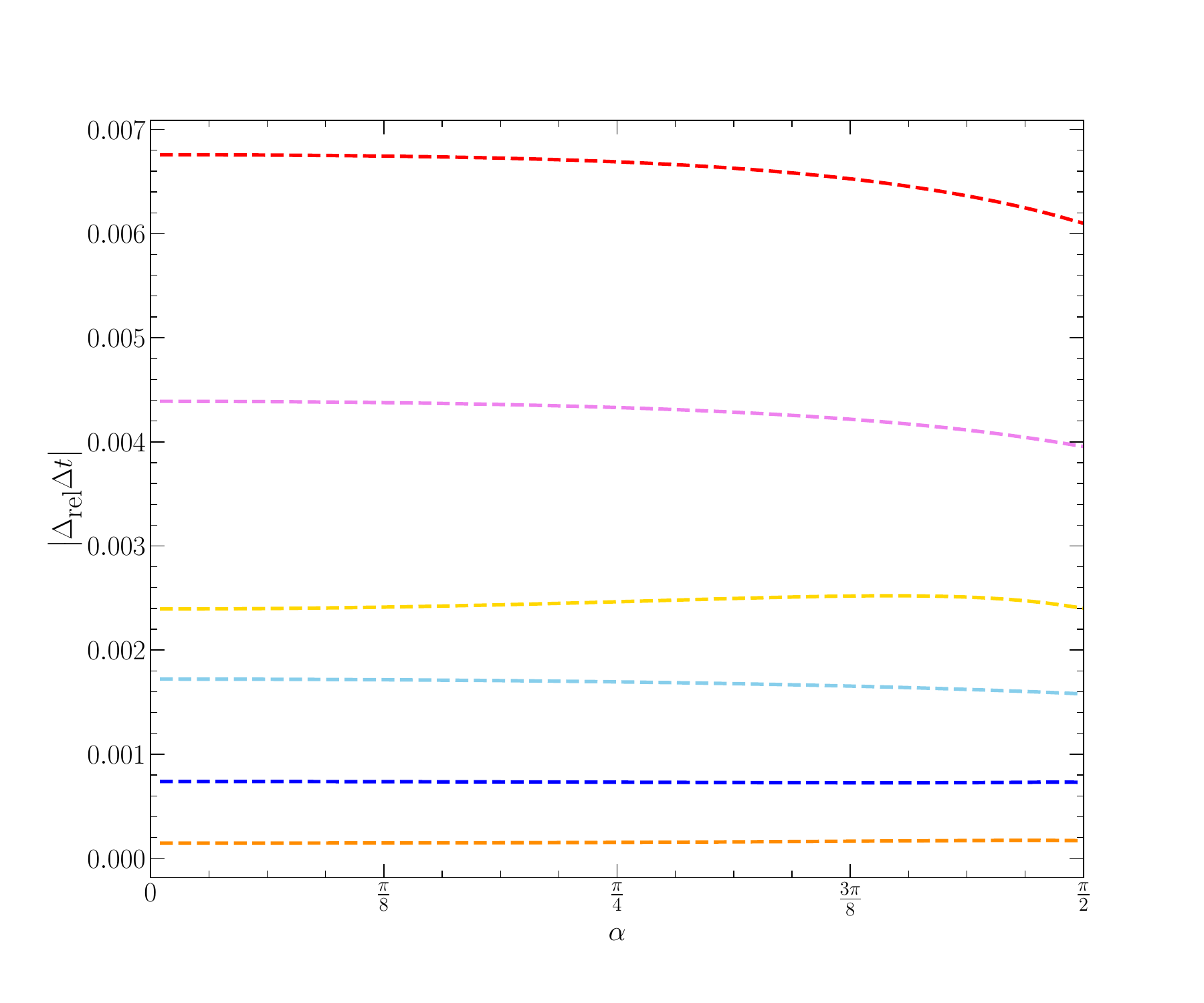}{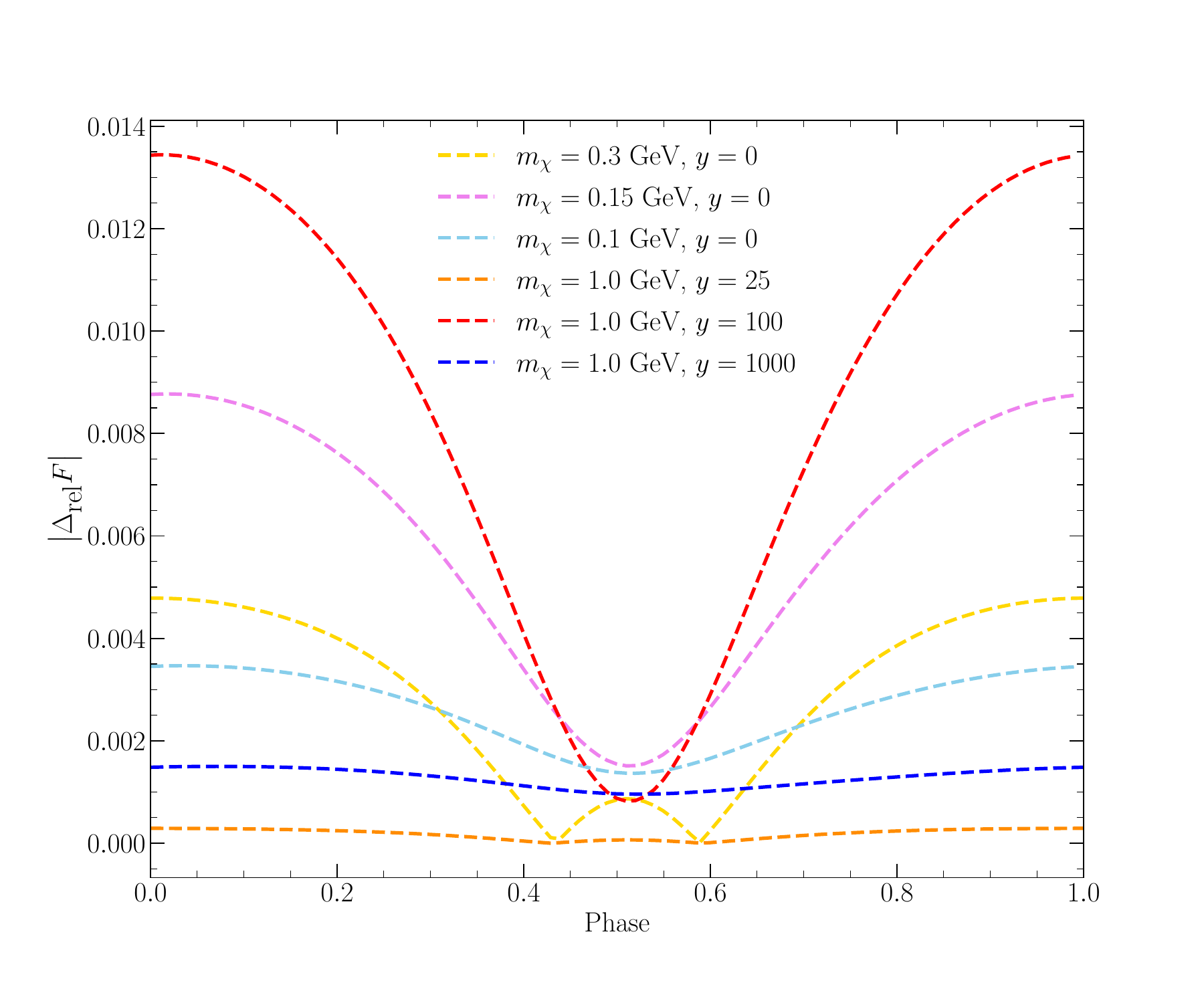}
\caption{Relative changes in gravitational self-lensing due to DM halos for DANSs with $M_{T} \left( R_{B} \right) = 1.33 M_{\odot}$ and $f_{\chi} = 0.05$. Changes are relative to the no-halo counterpart of each DANS with equal $R_{B}$ and mass $M_{T} = M_{T} \left( R_{B} \right) = 1.33 M_{\odot}$. Spin frequency and spot and observer geometries are the same as in Figure \ref{fig:F_vs_t}.
\label{fig:lens1.33}}
\end{figure}

\begin{figure}[ht!]
\plottwo{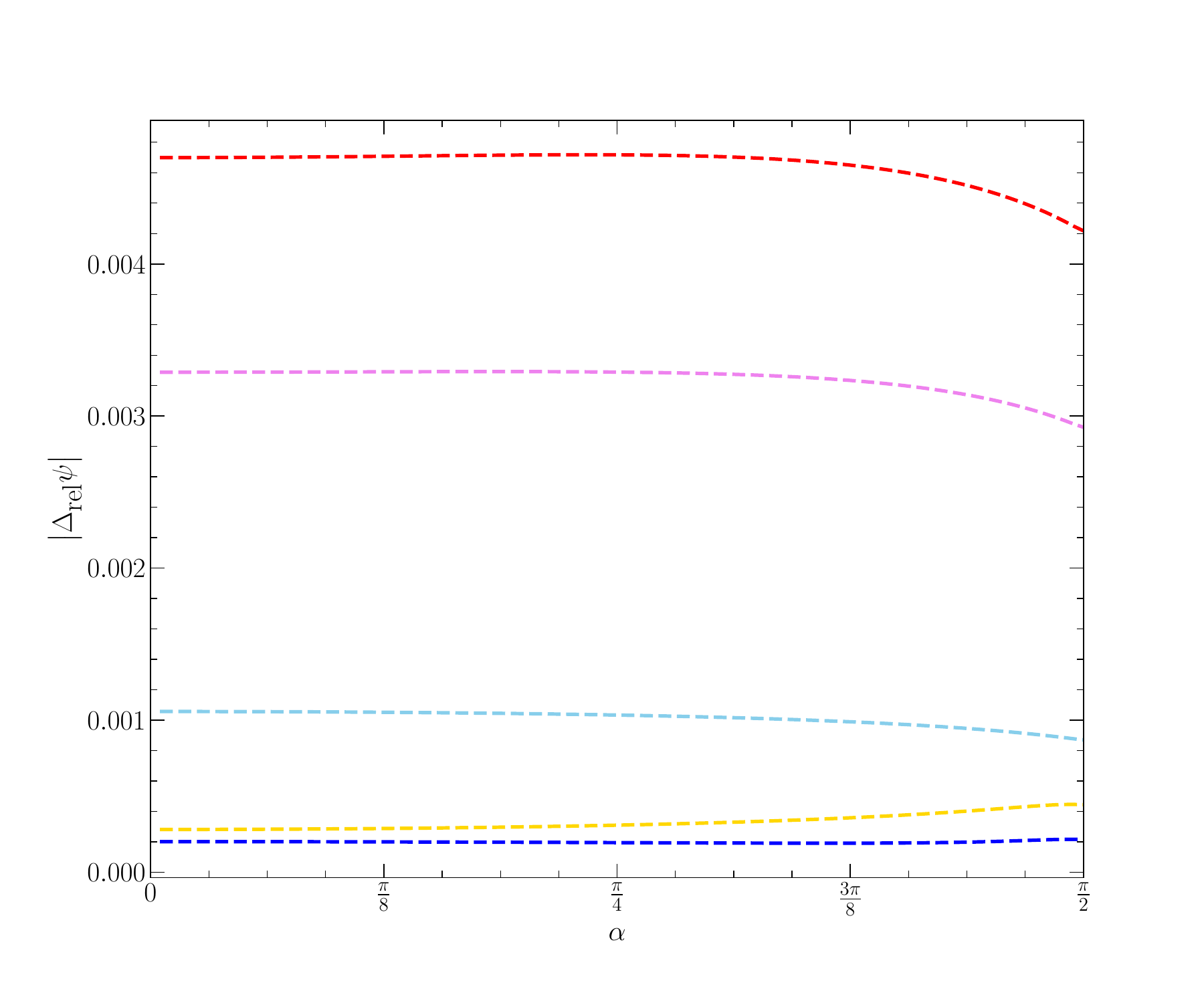}{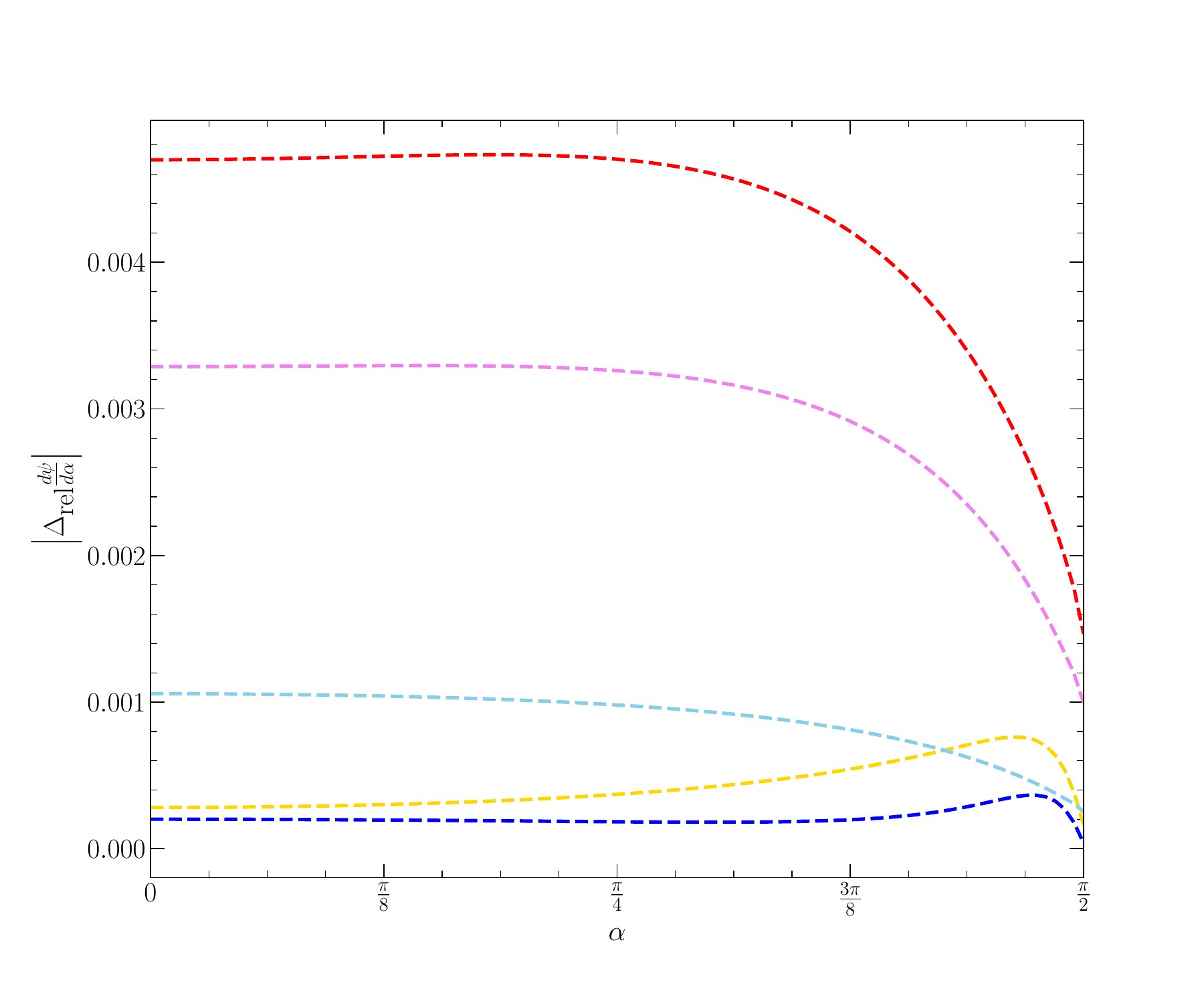} \plottwo{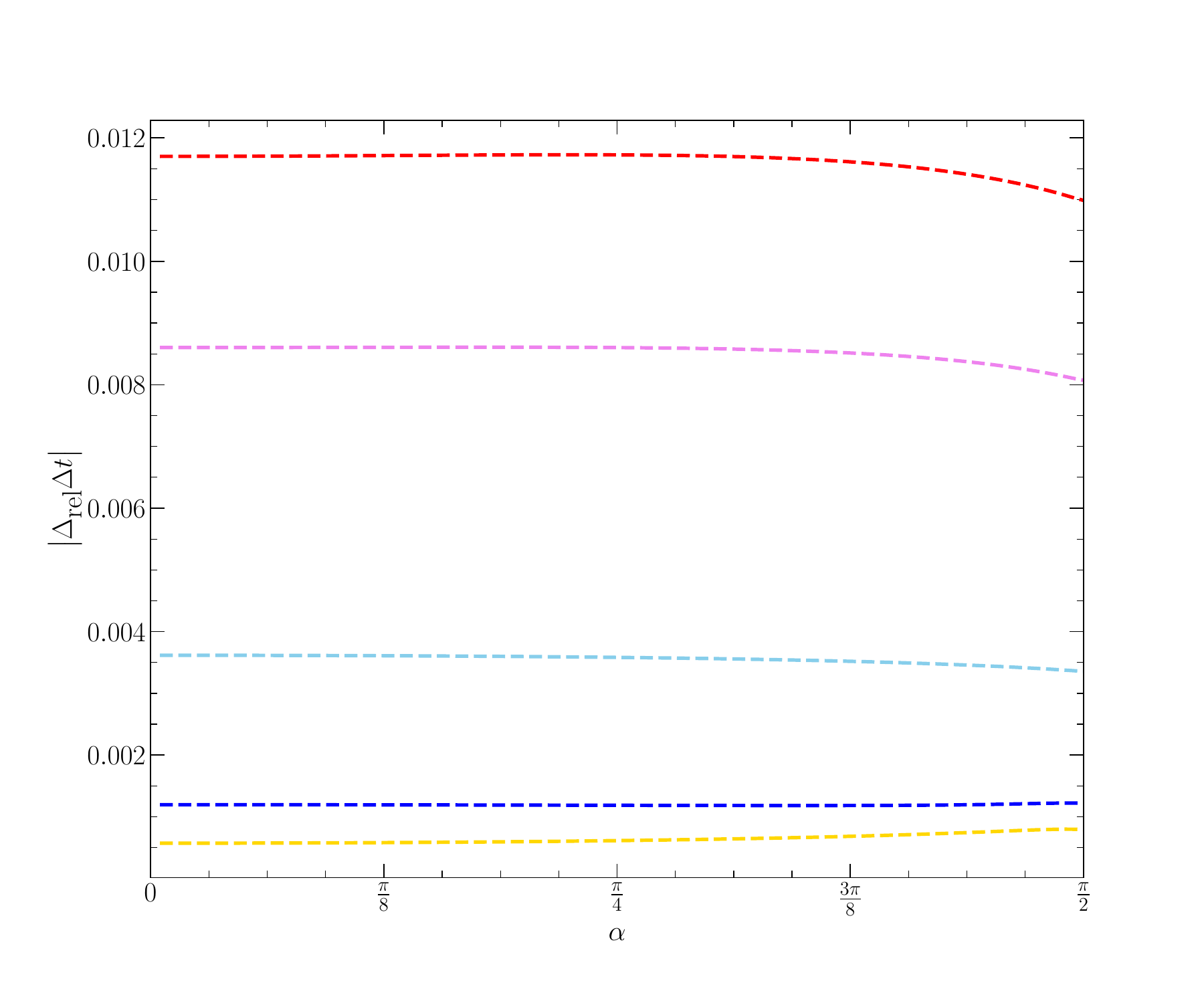}{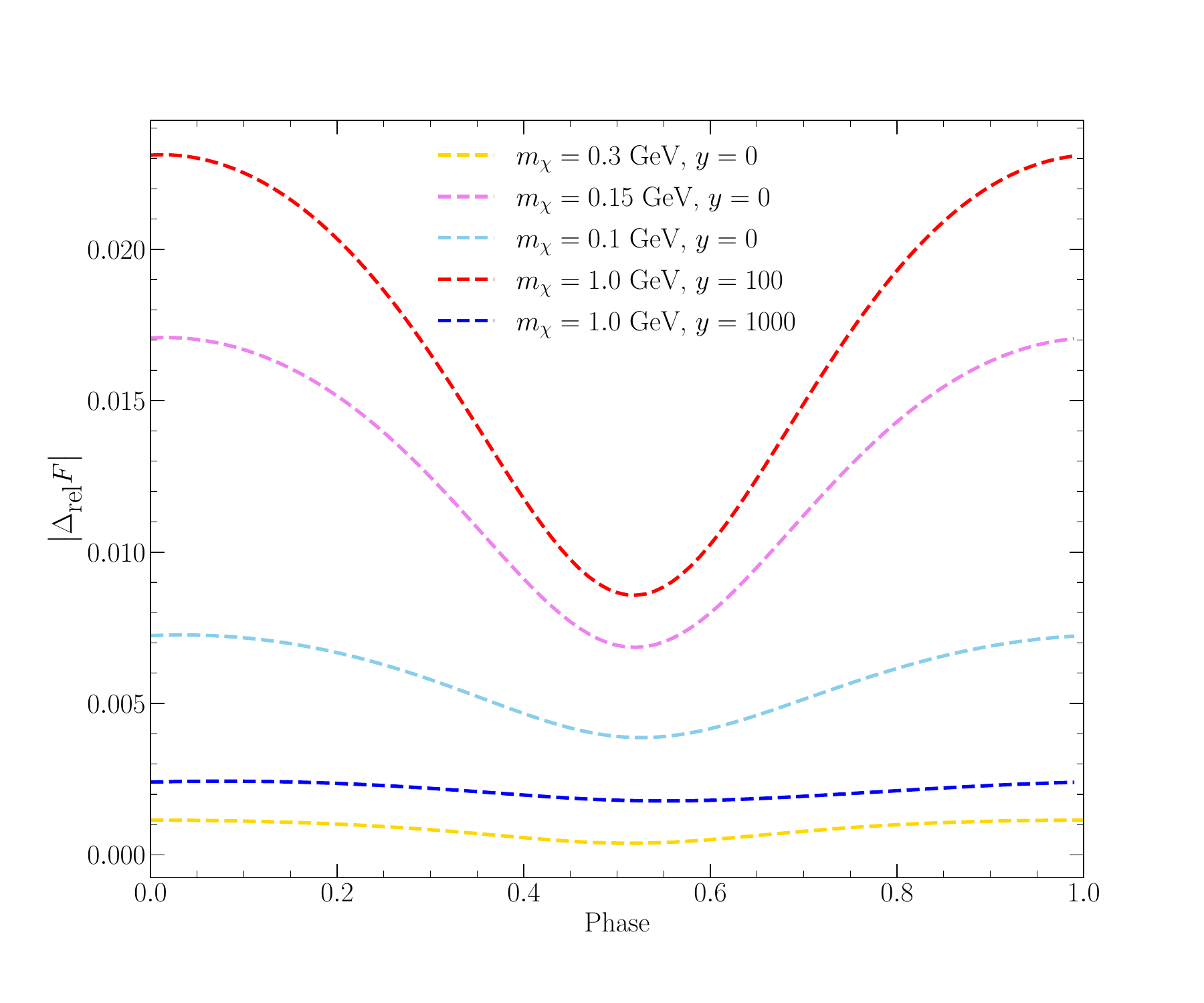}
\caption{Relative changes in gravitational self-lensing due to DM halos for DANSs with $M_{T} \left( R_{B} \right) = 2.00 M_{\odot}$ and $f_{\chi} = 0.05$. Changes are relative to the no-halo counterpart of each DANS with equal $R_{B}$ and mass $M_{T} = M_{T} \left( R_{B} \right) = 2.00 M_{\odot}$. Spin frequency and spot and observer geometries are the same as in Figure \ref{fig:F_vs_t}.
\label{fig:lens2.00}}
\end{figure}

\subsection{Comparison with the No-Halo Schwarzschild Counterpart NS}
\label{sec:NHSC}

The value  $e^{2\Phi}$, in the region between the baryonic surface and the outer edge of the DM halo is required in the calculation of the gravitational redshift, the acceleration due to gravity, and the gravitational light-bending. 
The metric function $\Phi$ can be simply computed by integrating Equation (\ref{eq:dphidr}), however, it is equivalent to rewrite the metric component in a way that refers to
the no-halo Schwarzschild counterpart introduced in the previous section. 
  Writing $e^{\Phi} = e^{-\Lambda + \Phi + \Lambda}$, and making use of Equation (\ref{eq:lambda}), $e^{\Phi \left( r \right)} = \left[ 1 - 2M_T \left( r \right)/r \right]^{1/2} e^{\Phi \left( r \right) + \Lambda \left( r \right)}$. Since $\Phi = - \Lambda$ in vacuum, the departure of $e^{\Phi+\Lambda}$ from unity provides a measure of the departure of the spacetime metric from the Schwarzschild metric due to the DM halo. In the region $r>R_B$, the sum of the two metric potentials can be found using Equation (\ref{eq:dphidr}) and the derivative of (\ref{eq:lambda})
and integrating from the edge of the dark halo inwards,
\begin{equation}
    \Phi(r) + \Lambda(r) = - 4\pi \int_r^{R_D} d\tilde{r} \frac{\tilde{r}^2}{\tilde{r} - 2M_T(\tilde{r})}\left[P_D(\tilde{r}) + \epsilon_D(\tilde{r})\right],
    \quad
    R_B \le r \le R_D.
    \label{eq:potsum}
\end{equation}
Equation (\ref{eq:potsum}) shows that the sum of the gravitational potentials is negative for a DANS with a halo, so that the tt-component of the metric is always smaller than the Schwarzschild value, and the gravitational redshift is always larger than the Schwarzschild value. 
The relative fractional difference $\Delta_{\textrm{rel}}$ between a quantity $Q$ at the baryonic surface and its Schwarzschild no-halo counterpart value, is defined by $\Delta_{\textrm{rel}} Q = (Q_{\textrm{Sch}} - Q)/Q_{\textrm{Sch}}$, where $Q_{\textrm{Sch}}$ is the value of Q evaluated at the same value of $r$ using the Schwarzschild metric with a value of mass equal to $M_T(r)$. The values of the fractional differences in the $g_{tt}$ component of the metric, $\Delta_{\textrm{rel}} g_{tt}(R_B)$ are shown for some representative models in Table \ref{tab:M_T(R_B)const}, and range from about $10^{-4}$ to close to $10^{-2}$. 

The change in the metric due to a dark matter halo is only one contribution to the change in the pulse profile, which also depends on the light-bending. To understand the contributions to the change in the pulse profiles due to changes in light-bending, it is useful to first consider the limit of small bending and emission angles.
In the limit of small $\alpha$, corresponding to photons emitted close to the normal to the surface, the integrals for the bending angle (\ref{eq:psi}), divergence of light rays (\ref{eq:dpsidalpha}), and the time delays (\ref{eq:deltat}) have the leading order values
\begin{eqnarray}
    \lim_{\alpha \rightarrow 0} \psi &=& \alpha \lim_{\alpha \rightarrow 0} \frac{d\psi}{d\alpha} =
    \alpha (1+z_B) J \label{eq:limpsi}\\
    \lim_{\alpha \rightarrow 0} \left[ \cos{\left( \alpha \right)} \frac{d \cos{\left( \alpha \right)}}{d \cos{\left( \psi \right)}} \right] &=& \left[ (1+z_B) J \right]^{-2} \label{eq:limdcosalpha}\\
    \lim_{\alpha \rightarrow 0} \Delta t &=& \frac{R_{B}}{2} (1+z_B)^2 \alpha^2 J \label{eq:limt}\\
    \lim_{\alpha \rightarrow 0} F &=& \gamma^{-5} (1+z_B)^{-4} J^{-2}, \label{eq:limF}
\end{eqnarray}
where the integral $J$, defined by
\begin{equation}
        J = \int_0^1 du \; e^{\Phi(u)+\Lambda(u)}
\end{equation}
is exactly equal to one if no DM lies outside the baryonic surface. In the case of a NS without a DM halo, all of these limits reduce to the respective limits for the no-halo Schwarzschild counterpart star with a mass and radius that coincide with the DANS's values of $M_T(R_B)$ and $R_B$, since in this limit $1+z_B = (1-2M_{T}/R_{B})^{-1/2}$. 

The relative differences $\Delta_{\text{rel}}$ for $\psi$, $d\psi/d\alpha$, and $\Delta t$ are plotted as a function of $\alpha$ for a few representative DM models in Figures (\ref{fig:lens1.33}) and (\ref{fig:lens2.00}). The largest relative differences typically occur close to $\alpha = 0$, so the limiting values shown in Equations (\ref{eq:limpsi}) - (\ref{eq:limF}) are most relevant. The limit of $\alpha = 0$ corresponds to the observed pulsar phase of $0$, so it can be seen from the plots of $\Delta_{\text{rel}}F$, that the largest relative changes in the bolometric flux also occur near $\alpha = 0$, where the flux is near its maximum.

The typical sizes of the relative changes in flux due to a DM halo can be understood by considering the size of the integral $J$. When a DM halo exists, Equation (\ref{eq:potsum}) ensures that the integrand $e^{\Phi+\Lambda}$ is less than or equal to one everywhere. In cases where the DM halo introduces small changes in the metric, we denote the small change by $\varepsilon$, defined by
\begin{equation}
    e^{(\Phi+\Lambda)_B} = 1 - \varepsilon/2, \varepsilon \ll 1. 
\end{equation}

The integrand of $J$ is equal to 1 in the region $0\le u \le R_B/R_D$, and can be roughly approximated by a straight line in the region $R_B/R_D \le u \le 1$, allowing the integral $J$ to be approximated by $J\sim 1 - \varepsilon (R_D-R_B)/(4R_D)$. 
In this limit of small $\varepsilon$, the maximum relative changes in the physical quantities (\ref{eq:limpsi}) - (\ref{eq:limF}) are, to first order in $\varepsilon$
\begin{eqnarray}
    \Delta_{\textrm{rel}} (1+z_B) &\sim & \frac12 \varepsilon\\
    \Delta_{\textrm{rel}} g_{tt} &\sim& - \varepsilon \\
    \Delta_{\textrm{rel}} F & \sim & - \frac12 \varepsilon \left( 3 + \frac{R_B}{R_D} \right) \\
    \Delta_{\textrm{rel}} dF_E & \sim & - \varepsilon \left( 1 + \frac{R_B}{2 R_D} \right).
\end{eqnarray}
In the case of very large DM halos the ratio $R_B/R_D$ is very small, while this ratio is close to one if the DM halo is similar in size to the baryonic surface. As a result, as long as the relative change in the metric is small, the relative change in the bolometric flux is roughly between 1.5 to 2 times larger than the change in the metric. Table \ref{tab:M_T(R_B)const} shows the relative changes in the metric and the bolometric flux for the representative DANS models. It can easily be seen that the maximum relative change in the bolometric flux is never larger than 2 times the relative change in the metric. Similar arguments hold for the relative change in the energy-dependent flux $dF_E$ which ranges from 1 to 1.5 times the relative change in the metric. As a result, if the relative change in the metric (at the baryonic surface) due to a DM halo is small, both the bolometric and energy-dependent flux will have small relative changes, and are well approximated by the no-halo Schwarzschild counterpart NS flux.
This is true for all types of halos.

\cite{Miao2022} calculated the magnitude of the maximum relative change in the flux
and found a empirical powerlaw relation between $\Delta_{\textrm{rel}} F$ and $M_{c}/R_{D}$. The ratio $M_{c}/R_{D}$ is half of the relative change in the metric evaluated at the outer edge of the dark matter halo. While our estimate of the maximum change in the flux depends on the change in the metric at the baryonic surface (instead of the dark matter surface) it doesn't matter which quantity is used in the estimation of the error in the flux, since if the difference in the metric is small at the baryonic surface, it will also be small at the dark matter surface.

\subsection{Dark Halos with Large Flux Deviation}
\label{sec:fluxdev}

The choice of how large the fractional error in the flux is allowed depends on the data. If the flux from the no-halo Schwarzschild counterpart is used to approximate the flux from the DANS, the largest fractional errors caused by the theoretical model should be smaller than the statistical errors. X-ray telescopes such as NICER count the number of photons in specific time and energy bins. The number count, $N_E$ is just the energy-dependent flux, Equation~(\ref{eq:dFE}), divided by the photon energy, so the fractional error in the number count scales the same way as $dF_E$. The statistical fractional error in the photon count is roughly $1/\sqrt{N_E}$. (In this simple estimate, we are not accounting for the background which can be a significant fraction of the signal.) In the case of PSR J0030, the largest number of photons in a time-energy bin is about 400 (see Figure 1 of \citet{Riley2019}), so the statistical fractional error is about 0.05, somewhat larger than the largest error introduced by the no-halo approximation for the intermediate halo case shown in red in Figure \ref{fig:lens2.00}. In other phase bins (corresponding to lower flux) the no-halo approximation has less error, while the statistical errors are larger. This suggests that using the no-halo approximation for the ``red" dark matter model is acceptable for modelling PSR J0030 at present, but in the future with longer exposures it might not be. The maximum fractional flux errors introduced for the other compact and diffuse halos are much smaller, so the no-halo approximation is accurate enough for these DANS models with the present observational data.

 When a dark halo is not present, the spacetime outside the baryonic surface of the star is approximated by the Schwarzschild metric in the SD approximation. The time-time component of the metric there has a simple analytical form.
 With the analytical form of the metric, the light-bending integrals presented in Section \ref{sec:SD}
are easily and quickly computed and used in the analysis of NICER data \citep{Bogdanov2019b}.

For DANSs with a halo, there is no analytical form for $\Phi(r)$ for $r < R_{D}$ so its value must be calculated numerically by solving Equations (\ref{eq:lambda} -- \ref{eq:dM_Ddr}). For dark halos that result in large flux deviations compared to the no-halo Schwarzschild counterpart (i.e. $\max{\left( \left| \Delta_{\textrm{rel}} F \right| \right)} > 1/\sqrt{N_{E}}$), the light-bending integrals must then be computed with this numerically calculated value of the metric for accurate results. 
It is not clear how to efficiently implement this in a Bayesian analysis.
In this case, one can not simply state that the gravitational-lensing observations have ``measured" $M_T(R_B)$ as can be done when the no-halo Schwarzschild counterpart is a good approximation. 
For these reasons, the extra computational cost makes this currently unfeasible for the halos that are poorly approximated by the no-halo Schwarzschild counterpart.

\subsection{Realistic Atmosphere Models}
\label{sec:atmos}

Realistic atmosphere models that predict the intensity of light from the surface of a NS depend on the acceleration due to gravity, $a$, also known as the surface gravity. In the region $r>R_B$ for a DANS with a halo, the surface gravity is $a = e^{-\Lambda} d\Phi/dr$, (for example \cite{2014AlGendy}) so the general expression for the surface gravity  is
\begin{equation}
    a = \frac{M_T(r)}{r^2 \left(1-2M_T(r)/r\right)^{1/2}} \left( 1 + 4 \pi r^3 \frac{P_D(r)}{M_T(r)} \right).
\end{equation}
The term proportional to the dark pressure is always very small, so the correction to the Schwarzschild expression for the surface gravity is tiny, as can be seen in Table \ref{tab:MRproperties} where the largest fractional changes are of order $1/1000$. This should be compared with an example of a realistic atmosphere model,  such as the Hydrogen atmosphere model nsatmos \citep{Heinke2006}. For nsatmos, a fractional change of at least 20\% in the surface gravity is required to change significantly the predicted flux. Clearly, the changes due to surface gravity introduced by a DM halo do not affect the emission of light from realistic atmospheres.

Atmosphere models for X-ray bursts on NS surfaces also depend on the ratio of the emitted flux to the Eddington limit \citep{2015A&A...581A..83Nattila}. The Eddington limit depends on the surface gravity, so using the no-halo Schwarzschild counterpart's value for the Eddington limit will be a good approximation for realistic atmospheres of X-ray bursters too. As a result, the mass appearing in the Eddington limit at the baryonic surface of a DANS will be $M_T(R_B)$.  

\subsection{Unpulsed Emission from the Surface of a DANS}
\label{sec:qlmxb}

Observations of unpulsed emission from the surface of a NS can also be used to constrain its mass and radius. Since the light emitted from the surface is lensed by the NS, the energy-dependent flux will be given by Equation (\ref{eq:dFE}), and integrated over the portion of the NS that is emitting. Observations of NSs in quiescent low-mass X-ray binaries \citep{Rutledge1999, Heinke2006,Steiner2018} and NSs exhibiting thermonuclear X-ray bursts
(e.g. \citet{2009ApJ...693.1775Ozel-EXO}, \citet{2017A&A...608A..31Nattila}), and UV emission from the surface of a rotation-powered pulsar \citep{2019MNRAS.490.5848Gonzalez} have been used to constrain the NS mass and radius. 

The determination of a NS's mass and radius is made by assuming the appropriate realistic atmosphere model for the specific intensity, $I'(E',\alpha')$, appearing in Equation (\ref{eq:dFE}) for different choices of mass and radius, integrating over the surface, and comparing the theoretical spectrum (\ref{eq:dFE}) with the observed spectrum obtained by a telescope such as Chandra. 
So far, most applications of this method assume that the entire surface emits with a homogeneous effective temperature, although it is known that unresolved hot spots could lead to biases \citep{2016ApJ...826..162Elshamouty}.
Since the atmosphere model and the lensed flux are well-approximated by the flux from the no-halo Schwarzschild counterpart NS, this leads to constraints on the values of $R_B$ and $M_T(R_B)$ when applied to a DANS.

\begin{deluxetable}{lccccccccccccc}[ht!]
  \tabletypesize{\scriptsize} 
  \tablecolumns{13}
  \tablewidth{0pt}
  \tablecaption{
    Properties of DANSs with dark halos and $f_{\chi} = 0.05$, and their no-halo counterparts. The first six DANSs are lower mass stars with $M_{T} \left( R_{B} \right) = 1.33 M_{\odot}$, and the last five are higher mass with $M_{T} \left( R_{B} \right) = 2.00 M_{\odot}$. Gravitational self-lensing properties of these stars are plotted in Figures \ref{fig:lens1.33} and \ref{fig:lens2.00}.%
    \label{tab:M_T(R_B)const}
  }
  \tablehead{%
    \colhead{} &
    \colhead{} &
    \colhead{$m_{\chi} $} &
    \colhead{$y$} &
    \colhead{$M_{T} $} &
    \colhead{$M_{\textrm{c}} $} &
    \colhead{$R_{B} $} &
    \colhead{$R_{D} $} &
    \colhead{$\left| \Delta_{\textrm{rel}} g_{tt} \left( R_{B} \right) \right|$} &
    \colhead{$\max{\left( \left| \Delta_{\textrm{rel}} F \right| \right)}$} &
    \colhead{$\left| \Delta_{\textrm{rel}}a \left( R_{B} \right) \right|$} &
    \colhead{$M_{\textrm{c}}/R_{D}$} &
    \colhead{$M_{\textrm{c}}/M_{D}$}
    \\
    \colhead{} &
    \colhead{} &
    \colhead{$\left[ \textrm{GeV} \right]$} &
    \colhead{} &
    \colhead{$\left[ M_{\odot} \right]$} &
    \colhead{$\left[ M_{\odot} \right]$} &
    \colhead{$\left[ \textrm{km} \right]$} &
    \colhead{$\left[ \textrm{km} \right]$} &
    \colhead{$\left[ 10^{-3} \right]$} &
    \colhead{$\left[ 10^{-3} \right]$} &
    \colhead{$\left[ 10^{-3} \right]$} &
    \colhead{$\left[ 10^{-3} \right]$} &
    \colhead{}
  }
  \startdata
    pure NS & black & $0$ & $0$ & $1.33$ & $0$ & $13.71$ & $0$ & $0$ & $0$ & $0$ & N/A & N/A \\
    compact halo & yellow & $0.3$ & $0$ & $1.34$ & $0.0094$ & $13.44$ & $19.86$ & $2.5$ & $4.8$ & $0.93$ & $0.70$ & $0.140$ \\
    diffuse halo & pink & $0.15$ & $0$ & $1.39$ & $0.0594$ & $13.67$ & $123.18$ & $5.8$ & $8.8$ & $0.69$ & $0.71$ & $0.855$ \\
    diffuse halo & sky blue & $0.1$ & $0$ & $1.40$ & $0.0678$ & $13.69$ & $400.38$ & $2.6$ & $3.5$ & $0.16$ & $0.25$ & $0.969$ \\
    compact halo & orange & $1$ & $25$ & $1.33$ & $0.0005$ & $13.39$ & $14.01$ & $0.15$ & $0.29$ & $0.070$ & $0.050$ & $0.007$ \\
    intermediate halo & red & $1$ & $100$ & $1.39$ & $0.0550$ & $13.64$ & $48.54$ & $8.2$ & $13$ & $1.3$ & $1.7$ & $0.794$ \\
    diffuse halo & blue & $1$ & $1000$ & $1.40$ & $0.0702$ & $13.71$ & $526.41$ & $1.2$ & $1.5$ & $0.023$ & $0.20$ & $0.997$ \\
    \hline
    pure NS & black & $0$ & $0$ & $2.00$ & $0$ & $14.06$ & $0$ & $0$ & $0$ & $0$ & N/A & N/A \\
    compact halo & yellow & $0.3$ & $0$ & $2.00$ & $0.0015$ & $13.69$ & $15.54$ & $0.58$ & $1.1$ & $0.25$ & $0.14$ & $0.015$ \\
    intermediate halo & pink & $0.15$ & $0$ & $2.07$ & $0.0725$ & $13.96$ & $87.28$ & $11$ & $17$ & $1.8$ & $1.2$ & $0.699$ \\
    diffuse halo & sky blue & $0.1$ & $0$ & $2.10$ & $0.0978$ & $14.04$ & $326.40$ & $5.1$ & $7.3$ & $0.46$ & $0.44$ & $0.932$ \\
    intermediate halo & red & $1$ & $100$ & $2.07$ & $0.0749$ & $13.97$ & $44.57$ & $14$ & $23$ & $2.6$ & $2.5$ & $0.721$ \\
    diffuse halo & blue & $1$ & $1000$ & $2.10$ & $0.1049$ & $14.06$ & $521.64$ & $2.0$ & $2.4$ & $0.048$ & $0.30$ & $0.996$
\enddata
\end{deluxetable}

\section{Proposed Analysis Methods} \label{sec:methods}


In Section \ref{sec:model}, we identified the following two different masses associated with NSs with a DM halo: the total mass contained within the dark matter radius, $M_T$, and the total mass contained within the baryonic matter radius, $M_T(R_B)$. These two masses can be similar or different depending on whether the halo is compact or diffuse. In addition, there are two different types of mass measurements for NSs: dynamical measurements, and gravitational self-lensing measurements. Inference of a DANS' properties then depends on the types of mass measurements available in addition to its halo type. In order to make clear the different types of analyses that are required for the different cases, we examine in more detail three types of measurement scenarios, in order of increasing complexity: only dynamical mass measurement available, only gravitational self-lensing measurement available, and both types of mass measurements available.

To illustrate how the analyses should be carried out in the different scenarios, we provide examples of hypothetical DANS observations in Figures \ref{fig:BM-EOSeffects} and \ref{fig:only_radio_or_NICERdiffuse}, where DANS(1) has both dynamical and gravitational lensing measurements available, and DANS(2) and DANS(3) only have gravitational lensing measurements available.

\subsection{Only Dynamical Mass Measurement Available}
\label{sec:onlyradio}

Dynamical mass measurements involve observations of the motion of the NS and/or the companion star within a binary system through radio pulse timing, gravitational radiation detections, optical observations of the companion, or a combination of the above. All dynamical mass measurements measure the total gravitational mass of the star $M_{T}$, regardless of the assumption of the existence of a dark matter halo.

A measurement of a large dynamical mass can rule out a BM-DM EOS combination for a particular mass fraction if the MR curve's maximum value of $M_T$ is smaller than the measured dynamical mass. This is the same way that mass measurements are used to constrain pure NSs \citep{Ozel2016}. The complication for DANS is that two different DANS could have different dark matter fractions, $f_\chi$, so making definitive EOS constraints is more difficult.  Given a pair of BM and DM EOSs and $f_\chi$, the method for deciding whether the EOS pair is allowed depends on whether the maximum mass DANS has a diffuse or compact halo, or a core.

If a DANS has a diffuse halo, then almost all its dark matter is located in the cloud outside of the baryonic surface and the maximum total mass is a fraction $f_\chi/(1-f_\chi)$ larger than the maximum allowed mass for pure NS constructed with the same BM EOS (as discussed in Section \ref{sec:MR}). In the case of a diffuse halo, any baryonic EOS that is allowed by the measurement of a high-mass NS will continue to be valid if a diffuse DM halo of any mass fraction is added to it (see NL3$\omega \rho$L55 curves in Figure \ref{fig:only_radio_or_NICERdiffuse} (left)). If the baryonic EOS is too soft to support a particularly high dynamically measured mass without dark matter, then adding a diffuse halo with a large enough DM mass fraction would allow the DANS to reach the high mass (see SLy5 curves in Figure \ref{fig:only_radio_or_NICERdiffuse} (left)).

For compact DM halos or DM cores, the maximum total mass is less than or equal to the maximum allowed mass for the baryonic EOS. The decrease in the maximum total mass depends on both the baryonic and dark EOS, so the analysis required for compact halos is the same as the analysis used to constrain dark cores (see \cite{Karkevandi2022, Shakeri2024, Mariani2024} for examples).

Dynamical mass measurements use Kepler's laws for point masses to determine the masses of the component stars, along with observations of other relativistic effects such as periastron precession, orbital decay due to gravitational radiation, and time delays caused by the gravitational potential of the companion \citep{Ozel2016}. The highest dynamical mass measured through radio observations (at present) is PSR J0740+6620 with a mass of $2.08\pm 0.07 M_\odot$, measured through observations of Shapiro delay \citep{Fonseca2021}. We assume the companion does not have an extended dark matter halo that would affect Shapiro delay, an assumption that could be examined in the future.


Similarly, gravitational radiation detections of a NS merging with another compact object allow the accurate determination of the chirp mass and total binary mass, which depend on $M_T$ for the DANS. For example, the secondary compact object in the gravitational wave merger event GW190814 has a mass of $2.59 \pm 0.08 M_\odot$ \citep{Abbott2020}. The nature of this object is unknown, but if it is a DANS, then comparison with the values of $M_T$ for different EOS is appropriate.


It should be remembered that rotation can also increase the maximum mass of a NS.
For example, PSR J0952-0607 spins rapidly, with a spin frequency of 709 Hz and an inferred mass of $2.35 \pm 0.17 M_\odot$ \citep{Romani2022} (which should be interpreted as a dynamical mass measurement). For this rapid rotation rate, the maximum mass allowed by a baryonic EOS will be larger by about 1 - 5\% \citep{Konstantinou2022} competing with any increase in the maximum value of $M_T$ arising from a diffuse dark matter halo.

\subsection{Only Gravitational Self-lensing Measurement Available}
\label{sec:onlyNICER}

For many NSs, X-ray emission from the surface is detected, but a dynamical mass measurement is not possible. Some of these are isolated NSs such as the rotation-powered X-ray pulsars  
PSRs J0030$+$0451 and J2124$-$3358 \citep{Bogdanov2019a} observed by NICER. Others are in binary systems, however, the observations of the binary motion do not presently provide constraints on the NS's mass. Examples include the rotation-powered X-ray pulsars  J1231$-$1411 \citep{Bogdanov2019a} and J0614$-$3329  \citep{Guillot2019}, and all of the non-pulsing NSs in quiescent low mass X-ray binaries \citep{Steiner2018}. In these cases, the light emitted from the surface is gravitationally self-lensed so that the equations in Section \ref{sec:lens} describe the detected flux. 

For most halos, the no-halo Schwarzschild counterpart NS approximates well the gravitationally-lensed flux. In these cases, the MR constraints arising the pulsed or unpulsed emission are constraints on the values of $R_B$ and $M_T(R_B)$. The decision about whether or not a MR constraint arising from an analysis of NICER data holds for a halo DANS depends on the statistics of the observation and the properties of the halos, as discussed in Section \ref{sec:NHSC}. The diffuse and compact halos are well-approximated by the no-halo Schwarzschild counterpart for small $f_\chi$. 
However, some intermediate halos can add enough 
 extra gravitational lensing that could lead to some systematic errors in the interpretation of their flux.
In Table~\ref{tab:M_T(R_B)const},
we show more examples of such intermediate halos with $M_{c}/M_{D} \sim 0.7 - 0.8$ and $f_{\chi} = 0.05$, and find that the observed X-ray flux from these DANSs have $>1\%$ deviation from that observed from no-halo Schwarzschild counterpart stars.

In cases where there is only a constraint on the NS mass and radius coming from gravitational self-lensing, a particular BM-EOS is valid if its $M_T(R_B)$ versus $R_B$ curve intersects the MR confidence region. The situation for a DM-EOS that produces diffuse halos is shown in 
Figure \ref{fig:only_radio_or_NICERdiffuse} (right). Since adding more dark matter to a DM-EOS that produces a diffuse halo leaves the $M_T(R_B)$ curve unchanged (for both stiff and soft BM-EOSs), if the measurement allows a particular pure BM-EOS, then any $f_\chi$ that produces a diffuse halo is also allowed (as long as the added DM does not violate the flux accuracy requirement). In the case of diffuse halos, the $M_T$ curves are irrelevant. In the hypothetical situation shown in Figure \ref{fig:only_radio_or_NICERdiffuse} (right), the observational constraints coming from DANS(2) allow the stiff EOS 
NL3$\omega \rho$L55 with any $f_\chi$ for a diffuse halo. Similarly, the soft EOS 
SLy5 is ruled by this observation, even if a diffuse halo is added.

Figure \ref{fig:BM-EOSeffects} (left) illustrates how compact halos can be constrained if only NICER MR measurements are available, as in the case of the cartoon measurement of fictitious DANS(3). The situation for compact halos is similar to how dark cores are constrained (as described by \cite{Rutherford2023}) since for both the MR curve becomes softer than the pure baryonic MR curve after the addition of DM. 
MR measurements are consistent with a BM-EOS if the 
 $M_{T} \left( R_{B} \right)$ and $R_{B}$ curve intersects with the NICER posterior, as in the example of  NL3$\omega \rho$L55 Figure \ref{fig:BM-EOSeffects} (left). Although the cartoon NICER observation of DANS(3) supports both stiff and soft pure BM-EOS, the addition of a $f_\chi=0.05$ compact DM halo is inconsistent with the soft EOS SLy5. 

For the DANSs in Figure \ref{fig:BM-EOSeffects} (left), $M_{T} \left( R_{B} \right) \approx M_{T}$. However, certain DM parameters can also produce compact halos whose $M_{T} \left( R_{B} \right)$ and $M_{T}$ can have larger differences such as the stars near the maximum mass of the $m_{\chi} = 0.15$ GeV, $y = 0$ (pink) curve in Figure \ref{fig:MR-yconst}. In such cases, it must be ensured that $M_{T} \left( R_{B} \right)$ is used for the analysis. 

\citet{Rutherford2023} restrict their proposed analysis to DANS without halos. However, the similarity of the compact halos' effect on $R_B$ means that their proposed analysis will also be valid for compact halos.

\subsection{Both gravitational self-lensing and dynamical mass measurements available}
\label{sec:bothNICERradio}

A few of the rotation-powered X-ray pulsars are in binary systems where the NS's mass can be dynamically determined, providing a measurement of $M_T$. If pulse-profile modelling of the pulsed emission can constrain the NS's mass and radius, this then allows the independent measurement of $R_B$ and $M_T(R_B)$. However, in practice, the parameter estimation analysis of the gravitationally-lensed X-ray emission is computationally expensive, so independent measurements of $M_T$ and $M_T(R_B)$ are not possible at present. Instead, when a dynamical mass measurement is available, the dynamical mass is used as a prior probability distribution in the Bayesian parameter estimation for the gravitational self-lensing mass and radius.
For example, in the Bayesian analysis of the NICER data for
J0740$+$6620 \citep{Riley2021, Miller2021} and PSR J0437$-$4715 \citep{Choudhury2024}, a 
Gaussian mass prior $p\left( M \right)$ is used, where
\begin{equation}
p \left( M \right) = \mathcal{N} \left( \mu_{M}, \sigma_{M} \right),
\label{eq:gaussian}
\end{equation}
where $\mathcal{N} \left( \mu_{M}, \sigma_{M} \right)$ is the Gaussian (normal) distribution function with mean $\mu_{M}$ and standard deviation $\sigma_{M}$ 
given by the radio observations. Other millisecond X-ray pulsars with dynamical mass measurements include
J0751$+$1807, J1012$+$5307 \citep{Guillot2019}, and J1614$-$2230 \citep{Wolff2021}.

Using a radio mass prior in the analysis of NICER data effectively assumes that the dynamical and gravitational lensing mass are the same, which is equivalent to assuming that $M_T$ and $M_T(R_B)$ are the same. This assumption is not a problem for the cases of DANSs with cores, since these two masses are the same.  However, a DANS with a diffuse halo has $M_T(R_B)$ that is reduced by a fractional difference $f_\chi$ from the total DANS mass, $M_T$. 

The situation for diffuse halos is illustrated schematically in 
Figure~\ref{fig:BM-EOSeffects} (right panel), for the hypothetical measurement of DANS(1). Radio observations provide the total mass, with a one-sigma confidence region shown as a green rectangle (since radio observations do not constrain radius). The Gaussian mass prior based on the radio observation is used in the analysis of the NICER X-ray data, leading to the one-sigma posterior distribution of the mass and radius shown as an orange oval.  Although the Gaussian mass prior allows for smaller and larger masses, the posterior distributions of the mass resulting from the X-ray data typically are very similar to the mass prior. We show this in the diagram as an ellipse with an average and range in the mass dimension that is identical to the radio prior. In this cartoon analysis that assumes $f_\chi = 0$, the stiff BM-EOS is supported by the data while the soft BM-EOS is invalid.

To correctly test for the value of $f_{\chi}$ in a DANS with a diffuse dark halo, $f_{\chi}$ needs to be added as an additional parameter in the Bayesian analysis of the NICER data, with $f_{\chi} \in \left[ 0, f_{\chi, \textrm{max}} \right]$ where $f_{\chi, \textrm{max}}$ is the maximum value of $f_{\chi}$  coming from other physically motivated considerations. With each tested value of $f_{\chi}$ between 0 and $f_{\chi, \textrm{max}}$ the mean of the Gaussian prior must be reduced by a factor of $f_{\chi}$,
\begin{equation}\label{eq:newpriormean}
p \left( M \right) = \mathcal{N} \left[ \mu_{M} \left( 1 - f_{\chi} \right), \sigma_{M} \right].
\end{equation}
The red ellipse in Figure \ref{fig:BM-EOSeffects} (right) shows the hypothetical posterior obtained testing for $f_{\chi} = 0.05$, assuming $M_{T} \left( R_{B} \right)/R_{B}$ stays constant between the original and new posteriors.
(It is impossible to know whether the median and one-sigma spread for the radius values will increase or decrease without redoing the full analysis.)
In such a scenario, any valid BM-EOS must have its $M_T(R_B)$ vs $R_B$ curve intersect the new posterior. Note that by construction, the values of $M_T$ for these DANSs will automatically agree with the dynamical mass measurement.

In Figure \ref{fig:BM-EOSeffects} (right), the pure BM-EOS NL3$\omega \rho$L55 was supported by the original posterior, but is invalidated by the new one. 
In the same figure, the situation for SLy5 was described in Section \ref{sec:onlyradio} when only dynamical mass measurement is available. This soft BM-EOS is left untested by the original posterior from gravitational lensing observations due to limited sampling in the low probability tail of the Gaussian mass priors. With the addition of the new posterior from gravitational lensing observations, SLy5 must intersect the red ellipse to be deemed valid. Thus, gravitational lensing observations added to the dynamical mass measurement are needed to assess whether a soft BM-EOS 
previously ruled out by a high mass measurement
can be revived with the addition of a diffuse halo.

The difference in the two masses is a problem in the example statistical analysis done by 
\citet{Miao2022} which makes use of the NICER observation of PSR~J0740 in an attempt to constrain the dark matter properties of DANS with $M_c/R_D < 10^{-3}$. The constraint on $M_c/R_D$ ensures that the X-ray flux is well-approximated by the no-halo Schwarzschild counterpart. However, both compact and diffuse halos with a wide range of $f_\chi$ are allowed by the $M_c/R_D$ requirement. Many of the DANSs sampled by \citet{Miao2022} are DANSs with diffuse halos, so the NICER mass and radius can't be used to constrain $M_T(R_B)$ and $R_B$ as is done in their analysis. Only the radio Shapiro delay mass measurement should be used as a constraint on $M_T$ for these cases. 

The problem with using the dynamical mass prior in a NICER analysis for diffuse halos is only an issue if the assumed dark matter mass fraction is large enough that the mass difference can be measured. For instance, if a maximum dark matter mass fraction of $f_\chi=0.05$ is assumed, there is a 5\% difference between $M_T$ and $M_T(R_B)$ for diffuse halos. This is similar to the size of the one-sigma range ($\sim 3\%$) in the dynamical mass measurement of PSR J0740 so the mass difference is in principle measurable, and a revised mass prior should be used. However, if a smaller dark mass fraction, for instance, $f_\chi = 10^{-3}$, were assumed, then the mass difference of 0.1\% for diffuse halos would be small enough that it would not be resolvable.

\subsection{Statistical EOS Inference}
\label{sec:inference}

The macroscopic properties of a DANS depend on both the EOSs of BM and DM. For a choice of $f_\chi=0.05$, the changes in mass and radius (compared to a pure NS) are on the order of at most 5\%. Meanwhile, the differences in mass and radius due to the BM-EOS for pure NSs can range to about 20\% (see for example, Figure \ref{fig:BM-EOSeffects}). Since the BM EOS is so important in determining a DANS's properties, constraints on DM properties arising from electromagnetic observations of DANS must account for the full range of possible baryonic EOS. For example, it would be incorrect to rule out the DM model with $m_\chi= 1$~GeV and $y=25$ using only the soft BM-EOS and the observation of DANS(3) in the example shown in Figure~\ref{fig:BM-EOSeffects}, since this dark matter model is compatible with the stiff BM-EOS. For this reason, one can't constrain DM properties by assuming only one BM-EOS (as is done in some cases, eg. \citet{Miao2022,Shakeri2024}).

Based on the complications introduced by DM, we imagine a statistical approach to constraining the EOS similar to the approach described by 
\citet{Miller2020}, but broadened to allow for dark matter, similar to the methods discussed by \citet{Rutherford2023}. Following their approaches one first introduces a set of parameters, $\alpha$ describing the baryonic EOS, and adds extra parameters describing the DM, such as $m_\chi$ and $y$. In this framework, there is one correct set $\alpha$ that describes all DANSs, however, each DANS could have a different value of $f_\chi$ and central baryonic and dark matter central densities. Physical considerations should be used to choose a maximum allowed $f_\chi$ for the statistical inference. 

A set of $n$ DANSs is observed, each with a separate posterior probability distribution for its mass and radius and the likelihood that $\alpha$ and a set of central densities and $f_\chi$'s describe each measurement. Before this statistical inference is performed, the type of measurement must be identified for each DANS. If the measurement is purely dynamical, for example the Shapiro delay observation of PSR J0740, the mass posterior corresponds to $M_T$. If the measurement only makes use of gravitational self-lensing, as, for example the NICER observation of PSR J0030, the mass and radius posteriors correspond to $M_T(R_B)$ and $R_B$, as long as the flux accuracy requirement is met (see Section \ref{sec:NHSC}). If a dynamical mass measurement was used as a prior in the gravitational lensing analysis (as is done in the NICER analysis for PSR J0740) then the gravitational lensing posterior distribution on the mass and radius can not be used to compute the likelihood for DANS with diffuse halos. In the case of diffuse halos, likelihoods for pure dynamical or pure gravitational-lensing observations will be identical to the likelihood for a pure NS without DM. Compact halo DANS can be treated in a manner similar to core DANS as described by \citet{Rutherford2023}.

\section{Conclusions} \label{sec:conclusions}

Adding dark matter to a NS effectively stiffens or softens the EOS depending on the dark matter's properties, complicating the inference of the EOS of cold, dense baryonic matter.  We, for the first time, have identified the pitfalls in interpreting the macroscopic properties of neutron stars with DM halos using radio and X-ray observations.

In Section \ref{sec:model}, we used the two-fluid TOV equations to simulate NSs admixed with fermionic DM with varying self-interaction strengths.
In the DANS models with a DM halo, DM exists both inside and outside the visible radius of the NS. A halo DANS has two different masses: $M_T$, the total mass (dark and baryonic) found within the outer radius of the DANS, and $M_T(R_B)$, the total mass inside $R_B$. We found it useful to classify halo DANS into two categories. 
In a compact halo, most of the DM is located inside the baryonic surface, which decreases $R_B$, $M_T$, and $M_T(R_B)$ compared to a pure baryonic NS. 

In a diffuse halo, most of the DM is located in a cloud outside the baryonic surface, and as a result, the values of $R_B$ and $M_T(R_B)$ are very close to the values for a pure baryonic NS. In a diffuse halo, the total mass of the DANS is 
larger than $M_T(R_B)$ with a fractional difference similar in size to the dark matter mass fraction $f_\chi$.

 We computed the mass distribution of the baryonic and dark components in each of these types of halos, and illustrated their effects on MR curves of constant $f_\chi$ in Section \ref{sec:MR}. In the case of DM producing a compact halo or a core for the maximum mass DANS, the curves of $M_T(R_B)$ or $M_T$ versus $R_B$ appear to overlap (at the resolution that we can measure), lying at smaller values of mass and radius than the pure baryonic MR curve (see Figure \ref{fig:BM-EOSeffects} (left panel)) leading to a softening of the EOS. On the other hand, if the maximum mass DANS has a diffuse halo, the $M_T(R_B)$ versus $R_B$ curves overlap the pure baryonic MR curves. The $M_T$ versus $R_B$ curve overlaps the baryonic curve for small masses and peels away from the baryonic curve to higher values of mass, as shown in Figure \ref{fig:BM-EOSeffects} (right panel). 
 
 We also investigated how the gravitational self-lensing properties of NSs change if dark halos exist in Section \ref{sec:lens} and showed how the no-halo Schwarzschild counterpart approximation (introduced by \citet{Miao2022}) arises.

Finally, in Section~\ref{sec:methods} we show how hypothetical electromagnetic measurements of the masses and radii of NSs with dark halos should be interpreted. 
Three possible types of observations of a DANS could be available: only a radio-timing observation; only X-ray emission from the surface; or both radio-timing and surface X-ray emission is observed. 
Since dynamical mass measurements constrain $M_T$ while gravitational self-lensing constrains $M_T(R_B)$ and $R_B$, special care is needed to make use of observations of pulsars such as PSR J0740 where a radio dynamical mass measurement is used as a prior for the statistical inference of the mass and radius from X-ray timing data obtained by NICER.

To implement this proposed framework for constraining dark halos with X-ray observations of NSs, we need to know whether a particular combination of BM and DM EOSs results in a DM diffuse or compact halo. In this work, we only included two different BM EOS as an initial exploration. We found that stiff and soft BM EOS have similar halo properties given the same DM parameters. Future work should explore the relation between BM and DM EOSs for a wider range of BM EOS and quantify the types of resulting halos given different parameters.

\begin{acknowledgments}
We thank Nathan Rutherford, Anna Watts, and the participants of the workshop INT-22R-2A ``Neutron Rich Matter on Heaven and Earth" and the ICDMS 2023 conference for useful conversations and valuable insights. We are grateful to the Institute for Nuclear Theory (INT) at the University of Washington for providing funding and a stimulating research environment during our visit. This research was supported in part by NSERC CGS-M and AGES scholarships awarded to S.S., by the INT's U.S. Department of Energy grant No. DE-FG02- 00ER41132, and NSERC Discovery Grant RGPIN-2019-06077.
\end{acknowledgments}

%

\vspace{5mm}








\bibliography{sample631}{}
\bibliographystyle{aasjournal}



\end{document}